\documentclass[twoside,leqno,twocolumn]{article}

\usepackage[letterpaper]{geometry}

\usepackage{ltexpprt}
\usepackage{hyperref}

\usepackage[noend]{algorithmic}
\usepackage{algorithm}
\usepackage{todonotes}
\usepackage{amsfonts, amsmath}
\usepackage{enumerate,enumitem}

\newcommand{\eps}{\varepsilon}
\newtheorem{definition}{Definition}
\newtheorem{observation}{Observation}

\begin{document}

\newcommand{\orc}{\includegraphics[height=\fontcharht\font`A]{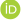}}

\newcommand\relatedversion{}
\renewcommand\relatedversion{\thanks{The full version of the paper can be accessed at \protect\url{}}} 

\date{}
\title{\Large Simple and Robust Dynamic Two-Dimensional Convex Hull}

\author{Emil Toftegaard Gæde\thanks{Technical University of Denmark, Kongens Lyngby, Denmark} \and Inge Li Gørtz \href{https://orcid.org/0000-0002-8322-4952}{\orc} \footnotemark[1] \and Ivor van der Hoog \href{https://orcid.org/0009-0006-2624-0231}{\orc}\footnotemark[1] \and  Christoffer Krogh\footnotemark[1] \and Eva Rotenberg \href{https://orcid.org/0000-0001-5853-7909}{\orc} \footnotemark[1]}

\maketitle







\begin{abstract}
The \emph{convex hull} of a data set $P$ 
is the smallest convex set that contains $P$. 
A \emph{dynamic} data set is one where points are inserted and deleted. In this work, we present a new data structure for convex hull, that allows for efficient dynamic updates, in theory and practice.

In a dynamic convex hull implementation, the following traits are desirable: (1) algorithms for efficiently answering queries as to whether a specified point is inside or outside the hull, (2) adhering to geometric robustness, and (3) algorithmic simplicity. 

Furthermore, a specific but well-motivated type of two-dimensional data is \emph{rank-based} data. Here, the input is a set of 
real-valued numbers $Y$
where for any number $y\in Y$ its rank is its index in $Y$'s sorted order. 
Each value in $Y$ can be mapped to a point $(\textsc{rank}, \textsc{value})$ to obtain a two-dimensional point set. Note that for a single update, a linear number of $(\textsc{rank}, \textsc{value})$-pairs may change; posing a challenge for dynamic algorithms. It is desirable for a dynamic convex hull implementation to also (4) accommodate rank-based data.

In this work, we give an efficient, \emph{geometrically robust}, dynamic convex hull algorithm, that \emph{facilitates queries} to whether a point is internal. Furthermore, our construction can be used to efficiently update the convex hull of \emph{rank-ordered data}, when the real-valued point set is subject to insertions and deletions. Our improved solution is based on an \emph{algorithmic simplification} of the classical convex hull data structure by Overmars and van Leeuwen~[STOC'80], combined with new algorithmic insights. 

Our theoretical guarantees on the update time match those of Overmars and van Leeuwen, namely $O(\log^2 |P|)$, while we allow a wider range of functionalities (including rank-based data). Our algorithmic simplification includes simplifying an 11-case check down to a 3-case check that can be written in 20 lines of easily readable C-code.  
We extend our solution to provide a trade-off between theoretical guarantees and the practical performance of our algorithm. 
We test and compare our solutions extensively on inputs that were generated randomly or adversarially, including benchmarking datasets from the literature.

\end{abstract}

\paragraph{Acknowledgements.}
This research was supported by Independent Research Fund Denmark grant 2020-2023 (9131-00044B) ``Dynamic Network Analysis'' and Eva Rotenberg's Carlsberg Foundation Young Researcher Fellowship CF21-0302 - ``Graph Algorithms with Geometric Applications''.
This project has additionally received funding from the European Union's Horizon 2020 research and innovation programme under the Marie Sk\l{}odowska-Curie grant agreement No 899987. 
\maketitle

\section{Introduction}\label{sec:introduction}
In analysis of spatial data, computing the convex hull yields one of the most fundamental characteristics of the data set. 
The convex hull itself serves as a computational stepping stone towards promptly answering relevant queries as they arrive online. Seen from the \emph{static} perspective, the problem of computing the convex hull has received much attention.
In this paper, we study the \emph{dynamic} setting, where data points may arbitrarily be added, deleted, or altered. 
Convex hulls have many applications for evolving or changing data sets, illustrating the need for an efficient, practical, and publicly available dynamic convex hull implementation.
%

\paragraph{Problem statement.}
The convex hull $CH(P)$ of a point set $P$ is the minimal convex shape that contains all points in $P$.
Convex hulls are  one
of the most prominent studied objects in computational geometry~\cite{brodal2002dynamic} due to their many applications which include top-queries~\cite{ihm2014approximate,mouratidis2017geometric}, clustering~\cite{liparulo2015fuzzy, gao2017trajectory, sander1998density}, road network analysis~\cite{liu2019fast, yan2011efficient, gao2017trajectory}, and data pruning/dimension reduction~\cite{giorginis2022fast, ostrouchov2005fastmap, khosravani2016convex, margineantu1997pruning}. 
A convex hull data structure can store $CH(P)$ either \emph{explicitly} or \emph{implicitly}.
An explicit data structure stores the vertices (or edges) of the hull, in their cyclical ordering, as a balanced binary tree. 

An  \emph{implicit} convex hull data structure  supports (a subset of) common convex hull queries without maintaining the edges of the convex hull.
The points of $P$ that lie on the convex hull of $P$ are called the \emph{extreme} points of $P$. 
An implicit data structure stores a data structure on $P$ that can answer queries on the convex hull such as~\cite{chan2011three}:
\begin{enumerate}[noitemsep]
    \item \label{it:extreme}
    Finding the extreme point of $P$ in a query direction, 
    \item Deciding whether a line intersects $CH(P)$, 
    \item Finding the two hull vertices tangent to a query, \label{it:tangent}
    \item Deciding whether a query point $q$ lies in $CH(P)$, \label{it:inside}
    \item Finding the intersection with a query line, \label{it:intersectline}
    \item Finding the intersection between two hulls. \label{it:hullintersect}
\end{enumerate}

We briefly note that with any explicit convex hull (storing $h$ points) one can answer each of these queries in $O(\log h)$ time.
We wish to dynamically maintain a convex hull supporting these queries subject to point insertions and deletions.

\paragraph{Related work.}
In the static setting, where the data set does not change, computing an explicit convex hull of a set of $n$ points
can be done in optimal $O(n \log n)$ time, e.g., with Graham’s scan~\cite{graham1972efficient} or Andrew’s vertical sweep line~\cite{andrew1979another}, or with the QuickHull $O(n \log n)$ expected time algorithm~\cite{barber1993quickhull}. 
Crucially, Graham's scan takes $O(n)$ time if the point set is already sorted by their radial ordering.
There also exist static, theoretically optimal, output-sensitive algorithms, due to Kirkpatrick and Seidel~\cite{kirkpatrick1986ultimate} and later Chan~\cite{chan1996optimal}:
They obtain $O(n \log h)$ running time where $h$ denotes the number of points of $P$ that lie on the convex hull. 
In a dynamic setting, the only data structure to explicitly maintain the convex hull is the $O(\log^2 n)$ algorithm by Overmars and van Leeuwen~\cite{overmars1980dynamically}.
This is also the best explicit dynamic convex hull algorithm with worst-case updates.

The first dynamic implicit data structure for the convex hull is by Chan~\cite{chan2001dynamic}, who achieves a linear-size data structure with $O(\log^{1 + \eps} n)$ amortized update time, and supports queries~\ref{it:extreme}--\ref{it:tangent} in $O(\log n)$ time (here, $\eps > 0$ is some arbitrarily small constant).
The update time was improved by Brodal and Jacob~\cite{brodal2002dynamic} to $O(\log n \log \log n)$ amortized. 
The original work by Chan~\cite{chan2001dynamic} can answer queries~\ref{it:inside}--\ref{it:hullintersect} in $O(\log^{3/2} n)$ worst case time.
This result was later improved by Chan~\cite{chan2011three} to support these queries in $O(\log^{1 + \eps} n)$ \emph{expected} time. 

\textbf{Convex hull implementations.}
Despite its numerous practical applications, implementations of dynamic convex hulls are scarce. 
Many approaches rely upon static algorithms for the convex hull (e.g.,~\cite{bennett2000geometry, wang2013online, bordes2005huller, ferragina2020pgm}), and thus have a linear update time. 
The CGAL library for computational geometry algorithms has no implementation of dynamic convex hull algorithms.
Instead, they use a dynamic algorithm for maintaining the Delaunay triangulation of a point set, which itself contains the convex hull. However, dynamic Delaunay triangulations have near-linear update time and linear recourse, 
rendering them useless for polylogarithmic dynamic updates. 
Chi, Hac{\'\i}g{\"u}m{\"u}{\c{s}}, Hsiung, and Naughton~\cite{chi2013distribution} provide a Java implementation of the dynamic algorithm by Overmars and van Leeuwen, which they use for an application in scheduling. Compared to ours, their algorithm lacks complete geometric robustness, it does not facilitate queries, and it cannot be extended to handle rank-ordered data.
Independently, Cisneros~\cite{Cisneros2007convex} presents a C implementation of the algorithm by Overmars and van Leeuwen.
Unfortunately, this implementation is insertion-only and has memory leaks. 
The algorithms for implicit convex hulls by Chan~\cite{chan2001dynamic} and Brodal and Jacobs~\cite{brodal2002dynamic} are not implemented and do not support the operations required for our applications.
The newer algorithm by Chan~\cite{chan2011three} does support these operations.
It is our impression that the three latter papers~\cite{chan2001dynamic,brodal2002dynamic,chan2011three}, 
present algorithms that 
are sufficiently complex that further simplifications and new ideas are needed if they should be implemented in a way that is efficient in practice.

\paragraph{Applications of (dynamic) convex hull.}
Convex hulls, or convex hull queries, have countless applications.
Here, we list some broad areas of application of dynamic convex hull queries. 
In Appendix~\ref{sec:applications} we elaborate on recent applications.

Query~\ref{it:inside} can be used for constraint satisfaction problems.
Indeed, consider a set of constraints given by a collection of halfplanes.
Each line supporting these halfplanes can be mapped to a point (i.e. its \emph{dual}).
The feasible region is subsequently given by the convex hulls of these points. Using Query~\ref{it:inside}, we can efficiently test if a candidate value satisfies all constraints. 

Query~\ref{it:intersectline} can, given a preference vector specifying the weight of the $x$ and $y$ coordinates, return the preferred element of $P$.
Repeated application of the convex hull can give the top $k$ preferred elements~\cite{mouratidis2017geometric, ihm2014approximate}.
Top-direction queries are often used as an intermediary pipeline step in database algorithms~\cite{chi2011icbs, chi2013distribution, qin2022mining}.

Query~\ref{it:hullintersect} can be used to determine a line that minimizes the maximal distance to $P$.
This has a direct application in (linear) regression, where we want to replace $P$ by a line that for any $x$-coordinate, can predict the corresponding value in $P$~\cite{chang2000onion, wilderjans2013chull, ferragina2020pgm}.

The explicit convex hull itself (represented as a subset of point $P' \subset P$) has applications as a pruning step for classification algorithms. 
Intuitively, the convex hull $P'$ is the subset of `extremal' points in $P$. 
Many classification algorithms are most sensitive to these extremal values. 
By maintaining the convex hull, one can subsequently run classification algorithms on smaller-complexity input~\cite{sander1998density, yuan2007convex, giorginis2022fast, jayaram2016convex}.
Similarly, sometimes only extremal points are interesting for learning algorithms (pruning the number of points in the training data). 
Examples include support vector machines (SVM) algorithms that use convex hull computations in their underlying algorithms~\cite{bennett2000geometry, wang2013online, chau2013convex, crisp1999geometric, mavroforakis2006novel}. Existing SVM algorithms use static convex hulls algorithms.  

We want to briefly note one special application of convex hull algorithms. 
Suppose that we are storing a (large) set of values $Y \subset \mathbb{R}$. 
The rank of each $y \in Y$, is its index in the sorted order of $Y$ and we want to support rank queries that for query values $q$ report the corresponding rank. If $Y$ is a set subject to insertions and deletions then this is the \emph{dynamic indexing} problem and it is well-studied~\cite{athanassoulis2014bf, bender2000cache, chan1998bitmap, pagh2004cuckoo, rao1999cache}. 
Recently, a new family of indexing data structures, called learned
indexes, has been introduced~\cite{kraska2018case, ferragina2020learned, ferragina2020pgm}. 
Consider a set of values $Y$ and the corresponding two-dimensional point set $P_Y$ that maps every value in $x$ to the coordinate $(\textsc{rank}, \textsc{value})$.
By maintaining the convex hull of $P_Y$, recent work~\cite{kraska2018case, ferragina2020learned, ferragina2020pgm} uses ML techniques that, for some given parameter $\eps$, predict the rank of each element with some additive error $\eps$.


\paragraph{Our contribution.}
Via new theoretical insights, we provide a new practical algorithm 
for the dynamic convex hull problem, and provide an efficient, geometrically robust implementation. Our implementation makes use of the CGAL CORE library,
to facilitate usage in geometric computations.
More specifically:

\begin{itemize}[noitemsep]
    \item We establish a characterization for where to insert edges into a dynamic convex hull based on edge-to-edge comparison. This allows for storing the convex hull as a collection of edges (as opposed to points) which simplifies update logic compared to~\cite{overmars1980dynamically, chi2013distribution}.
    \item We apply this characterization to the algorithm by Overmars and van Leeuwen~\cite{overmars1980dynamically} to get an implementation of a linear-size data structure to explicitly maintain a convex hull of size $h$ in a balanced binary tree with $O(\log^2 n)$ update time that facilitates queries in $O(\log h)$ time. 
    \item We observe that OvL~\cite{overmars1980dynamically}, even when simplified using our streamlined edge-based logic, comes with considerable overhead by maintaining what is called concatenable queues. 
    We propose a new definition of explicit convex hull, where we do not store the edges of the convex hull in their cyclical ordering. Instead, we only require that we may (contrary to previous implicit convex hulls) answer queries~\ref{it:extreme}--\ref{it:hullintersect} and that we can report the $h$ points on the convex hull in $O(h \log n)$ time. 
    \item With our edge-based logic, we construct a new algorithm, \textbf{Eilice}, for maintaining this hull in $O(\log^2 n)$ worst-case update time with $O(\log n)$ queries.  
    \item We provide both a geometrically robust version of Eilice, as well as an approximate version since both versions have applications in practice.
    \item For rank-based convex hulls, we present and implement new algorithms with the guarantees. 
    
    Note: These are the \underline{first dynamic algorithms} for rank-based hulls with 
    polylogarithmic update time.
    \item We compare our implementations (the simplified OvL and the new Eilice) to the preexisting dynamic convex hull implementations on a variety of generated and adversarial outputs. We conclude that: \begin{itemize}
        \item geometric robustness comes at an expense of computation time,
        \item as soon as the access pattern has less than a few hundred queries per update, dynamic algorithms have a clear advantage,
        \item we obtain improved performance compared to the state-of-the-art (including \cite{chi2011icbs,chi2013distribution}).
    \end{itemize} 
\end{itemize}

\begin{figure}[t]
\centering
\includegraphics[]{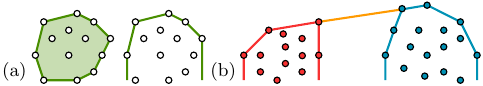}
\caption{
(a) The hulls $CH(P)$ and $CH^+(P)$. 
(b) Given two upper convex hulls $CH^+(R)$ and $CH^+(B)$, we construct their bridge.
}
\label{fig:convexhull}
\end{figure}

\section{Preliminaries}

Let $P$ be a set of $n$ points in $\mathbb{R}^2$. 
For any two points $(\alpha, \beta)$ we denote by $\overline{\alpha \beta}$ the segment between them. 
We say that a line $\ell$ separates $P$ whenever it contains at least one point of $P$ in the interior of either halfplane bounded by the line.
The convex hull $CH(P)$ is the edge-set of the minimal convex shape that contains all points in $P$. Equivalently, the convex hull consists of all points $P'$ such that each line through $(p_1, p_2) \in P' \times P'$ does not separate $P$.

We say a point $a$ \emph{precedes} a convex hull edge $\alpha$ if it has an $x$-coordinate that is lesser or equal than the left endpoint of $\alpha$.
The \emph{upper convex hull} $CH^+(P)$ is defined as the convex hull of the set $P \cup \{0, - \infty \}$ ((Figure~\ref{fig:convexhull} (a)).
The \emph{lower convex hull} $CH^-(P)$ is defined as the convex hull of $P \cup \{0, \infty \}$, and when we view convex hulls as an area in $\mathbb{R}^2$, $CH(P) = CH^+(P) \cap CH^-(P)$.

Consider two point sets $A$ and $B$ that are separated by a vertical line. 
The convex hull $CH(A \cup B)$ is formed by two segments of $CH(A)$ and $CH(B)$, together with up to two edges called \emph{bridges}. Commonly, a bridge is defined as a \emph{minimal} segment $\overline{ab}$ for points $(a, b) \in A \times B$ where the line extending $\overline{ab}$ does not separate $A \cup B$, see Figure~\ref{fig:convexhull}(b).
One of the bridges coincides with $CH^+(A \cup B)$, and one with $CH^-(A \cup B)$.

\paragraph{Partial Hull Trees}
Overmars and van Leeuwen~\cite{overmars1980dynamically} maintain the convex hull of a two-dimensional point set $P$ by maintaining $CH^+(P)$ and $CH^-(P)$ in two separate linear-size data structures that can answer each of the queries (1)--(6) in $O(\log h)$ time where $h$ is the size of the convex hull. 
By computing the points of intersection between $CH^+(P)$ and $CH^-(P)$ they can identify all edges of $CH(P)$.

They store  $CH^+(P)$ and $CH^-(P)$ in a data structure that we will refer to as a Partial Hull Tree (PHT). 
We recall the data structure for storing the upper convex hull $CH^+(P)$ as these data structures are symmetrical.

\begin{definition}
Given a two-dimensional point set $P$, the \emph{Partial Hull Tree} stores $P$ in a leaf-based balanced binary tree $T^+$ (sorted by $x$-coordinates).
For each interior node $v \in T^+$, denote by $\pi(v)$ the points in the leaves at the subtree rooted at $v$. 
For each node $v \in T^+$ with children $(x, y)$ and parent $v'$, the \emph{Partial Hull Tree} stores:

\begin{itemize}[noitemsep]
    \item  The unique bridge $e^+(v)$ between the upper hulls $CH^+(\pi(x))$ and $CH^+(\pi(y))$.
\item  A \emph{concatenable queue} $\mathbb{E}^*(v)$. 
This is a balanced binary tree of the vertices of $CH^+(\pi(v))$ that are not in $CH^+(\pi(v'))$ (where $v'$ is the parent of $v$). 
\item For the root $r$ of $T^*$, $\mathbb{E}^*(r)$ equals $CH^+(P)$.
\end{itemize}
\end{definition}

\begin{figure}[t]
\centering
\includegraphics[]{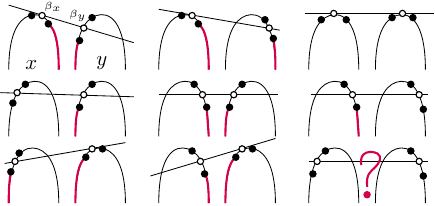}
\caption{The 
$9$ main cases for a line through the pivots $(\beta_x, \beta_y)$. For each case, \cite{overmars1980dynamically} can exclude at least $\frac{1}{4}$ of the remaining convex hull points as an endpoint of $e^+(v)$. The $9$'th case is further split into three others. 
}
\label{fig:elevencases}
\end{figure}

\paragraph{Dynamically maintaining a Partial Hull Tree.}
Let $p$ be a point added to or removed from $P$ and denote by $\rho$ the root-to-leaf path to $p$. 
In \cite{overmars1980dynamically}, they restore the Partial Hull Tree by computing for every node $v \in \rho$ the bridge $e^+(v)$. 
Let $x$ and $y$ be the children of $v$.
We define a \emph{pivot} $(\alpha_x, \beta_x, \gamma_x)$ as any triple of consecutive points on $CH^+(\pi(x))$. 
Overmars and van Leeuwen consider a pair of pivots $(\alpha_x, \beta_x, \gamma_x)$ and $(\alpha_y, \beta_y, \gamma_y)$ of $x$ and $y$.
Through a case distinction of $11$ cases (Figure~\ref{fig:elevencases}) they can conclude if $e^+(v)$ has its left endpoint preceding or succeeding $\beta_x$.
This gives the following dynamic update algorithm~\cite{overmars1980dynamically} that has two main steps:

\textbf{Traversing down. } First, they traverse the path from the root to the leaf containing $p$. For every node $v$ they encounter, they consider the concatenable queues $\mathbb{E}^*(v), \mathbb{E}^*(x), \mathbb{E}^*(y)$. Here $x$ and $y$ denote the left and right child of $v$, respectively. 
Through the split and join operations of balanced binary trees, they obtain the convex hulls $CH^+(\pi(x) \backslash p)$ and $CH^+(\pi(y) \backslash p)$ in $O(\log n)$ time as balanced binary trees $\mathbb{E}(x)$ and $\mathbb{E}(y)$. 

\textbf{Bubbling up. } Next, they traverse the path from the leaf containing $p$ to the root bottom-up and compute for every vertex $v$ on the path the bridge $e^+(v)$.
Let $v$ have children $x$ and $y$:

\begin{enumerate}[noitemsep]
        \item Each inner node of $\mathbb{E}(x)$ represents an vertex $\beta_x$. In $O(1)$ time they obtain the predecessor or successor of $\beta_x$ on $CH^+(\pi(x) \backslash p)$. The endpoints of these edges give a pivot $(\alpha_x, \beta_x, \gamma_x)$.
    \item Similarly, each inner node of $\mathbb{E}(y)$ comes with a pivot. Given two pivots, they decide in $O(1)$ time if the endpoints of $e^+(v)$ succeed, precede or coincide with the pivots (Figure~\ref{fig:elevencases}). Depending on the case, they recurse down $\mathbb{E}(x)$ and/or $\mathbb{E}(y)$.
    \item Since there are at most $O(n)$ points on $CH^+(\pi(x))$ and $CH^+(\pi(v))$, they recurse at most $O(\log n)$ times before they find the bridge $e^+(v)$ between $CH^+(\pi(x))$ and $CH^+(\pi(y))$.  
\end{enumerate}

Given the bridge $e^+(v)$, they update $\mathbb{E}^*(v)$. 
After having updated $\mathbb{E}^*(r)$ for the root $r$, they have computed the convex hull $CH^+(P)$.

\paragraph{Runtime analysis.}
There are $O(\log n)$ nodes $v$ on the root-to-leaf path $\rho$ in $T^+$. 
They iterate over $\rho$ to obtain for all $v \in \rho$ a binary tree $\mathbb{E}(v)$ in $O(\log^2 n)$ total time. 
Then, they traverse $\rho$ from $p$ towards the root.
For each $v \in \rho$ that the three steps are performed in $O(\log n)$ time.  Thus the total running time is $O(\log^2 n)$. 

\paragraph{The logarithmic method}
We briefly mention the logarithmic method, since Ferragina and Vinciguerra~\cite{ferragina2020pgm} used it to maintain convex hulls to predict the rank of elements in a fully dynamic set. 
The logarithmic method transforms (decomposable) static data structures into insertion-only dynamic structures (see e.g.~\cite{overmars1983design}).
It uses a bucketing scheme of buckets that increases in size exponentially.
In Appendix~\ref{app:logmethod} we describe the logarithmic technique applied to the convex hull, and why doing so leads to incorrect query algorithms and hulls.

\section{Simplified Algorithm for Convex Hulls}
\label{sec:with-cqueue}

\begin{figure}[h]
\centering
\includegraphics[page=2]{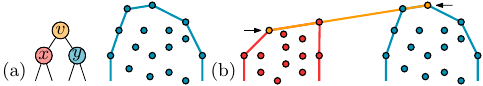}
\caption{ 
(a) $v \in T^+$ with children $x$ and $y$. We show the upper convex hull  $CH^+(\pi(y))$.
(b) We redefine the bridge $e^+(v)$ as the \textbf{maximal}  non-separating segment between $CH^+(\pi(x))$ and $CH^+(\pi(y))$.
}
\label{fig:configurations}
\end{figure}

In this section, we establish a new characterization for where to insert new edges into a dynamic convex hull.
In this analysis, we use edges of the current convex hull as pivots, as opposed to the analysis in~\cite{overmars1980dynamically} that uses points (Figure~\ref{fig:elevencases}). 
We then use it to provide a more straightforward implementation of the algorithm in~\cite{overmars1980dynamically}. 
Our analysis makes the algorithm in~\cite{overmars1980dynamically} not only simpler but also more efficient as we may store and traverse fewer pointers.
To this end, we denote by $T^+$ the Partial Hull Tree of $P$ where we \textbf{redefine} {bridges} and {concatenable queues}:

\begin{definition}
    \label{def:bridge}
    Let $v \in T^+$ have children $x$ and $y$. 
    We define its \emph{bridge} $e^+(v)$ as
a \textbf{maximal} segment $\overline{ab}$ for points $(a, b) \in CH^+(\pi(x)) \times CH^+(\pi(y))$ where the line extending $\overline{ab}$ does not separate $\pi(x) \cup \pi(y)$ (Figure~\ref{fig:configurations}).
\end{definition}

\begin{figure}[t]
\centering
\includegraphics[]{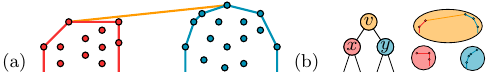}
\caption{
(a) Hulls $\mathbb{E}(x)$ and $\mathbb{E}(y)$. 
(b) The root $v$ stores $\mathbb{E}^*(v) = \mathbb{E}(v)$.
Its left child $x$ stores $\mathbb{E}^*(x) = \mathbb{E}(x) \backslash \mathbb{E}(v)$. 
}
\label{fig:cqueue}
\end{figure}

\begin{definition}
    Let $v \in T^+$ have children $x$ and $y$. 
    Denote by $\mathbb{E}(v)$ the edges of $CH^+(\pi(v))$ in their cyclical ordering. 
    We define $\mathbb{E}^*(x) := \mathbb{E}(x) \backslash \mathbb{E}(v)$ (Figure~\ref{fig:cqueue}). 
    The \emph{concatenable} queue (or c-queue) is $\mathbb{E}^*(x)$ stored in its cyclical ordering as a balanced binary tree. 
\end{definition}

\begin{lemma}
    Let $v \in T^+$ have children $x$ and $y$. 
    Given $\mathbb{E}(v)$, $\mathbb{E}^*(x)$, and $\mathbb{E}^*(y)$,  we may obtain $\mathbb{E}(x)$ in $O(\log n)$ time using the split and join operations on binary trees.
\end{lemma}

\begin{proof}
    The edges of $\mathbb{E}^*(x)$ are a contiguous interval in $\mathbb{E}(v)$. 
    Moreover, an edge is in $\mathbb{E}(x) \cap \mathbb{E}(y)$ if and only if it precedes the bridge $e^+(v)$. Thus, we achieve the lemma through  Algorithm~\ref{alg:hullget}. 
\end{proof}

Whenever we invoke Algorithm~\ref{alg:hullget}, we store the tree $\mathbb{E}_R$ so that we may invert the algorithm to obtain $\mathbb{E}(v)$ from $\mathbb{E}^*(x)$ in $O(\log n)$ time. \\

\paragraph{The Partial Hull Tree and its implementation.}
The Partial Hull Tree $T^+$ is a balanced binary tree on $P$ sorted by $x$-coordinate. 
Each node $v \in T^+$ with children $x$ and $y$ stores the bridge $e^+(v)$ between $CH^+(\pi(x))$ and $CH^+(\pi(y))$ and the c-queue $\mathbb{E}^*(v)$.
The Partial Hull Tree $T^-$ is defined analogously, using lower convex hulls.

We note that we can avoid storing $P$ twice: by maintaining only one tree $T$ where every node $v$ maintains both $e^+(v)$ and $e^-(v)$.
We choose to implement the Partial Hull Tree and all c-queues as AVL trees because we want to have worst-case guarantees.  
To simplify our internal algorithm's logic, we define the leaves of our $c$-queue to be points that are the endpoints of their parent edges. 
With slight abuse of notation, we refer indistinguishably between edges $\sigma \in \mathbb{E}(v)$ and their $c$-queue node. 
That is, for each node $w \in T$ there is a unique ancestor $v$ of $w$ where the bridge $e^+(w) = \sigma \in \mathbb{E}^*(v)$.

 \begin{algorithm}[h]
    \caption{splitHull(left child $x$, balanced tree $\mathbb{E}(v)$)}
    \label{alg:hullget}
    \begin{algorithmic}
    \STATE $(\mathbb{E}_L, \mathbb{E}_R  \gets \mathbb{E}(v)$.Split($e^+(v)$)
    \STATE $\mathbb{E}(x) \gets E_L(v)$.Join($\mathbb{E}^*(x)$)
    \RETURN $(\mathbb{E}(x), \mathbb{E}_R)$
    \end{algorithmic}
  \end{algorithm}

\subsection{Recomputing bridges using only edges\newline}

Let $v \in T$ have children $x$ and $y$. 
Our new definition of bridges and c-queues allows for new logic to recompute the bridge $e^+(v)$ from $\mathbb{E}(x)$ and $\mathbb{E}(y)$ in $O(\log n)$ time.
We show this through two lemmas: 

In Lemma~\ref{lem:decision}, we consider two edges $(\alpha, \beta) \in \mathbb{E}(x) \times \mathbb{E}(y)$.
In our data structure, these are both roots of a subtree in their respective trees.
We can decide in $O(1)$ time whether $e^+(v)$'s endpoints succeed or precede these edges: discarding at least one child of either $\alpha$ or $\beta$ until we find one endpoint of $e^+(v)$. 

In Lemma~\ref{lem:leaf}, let $e^+(v) = (a, b)$. Given $a$ and $\beta \in \mathbb{E}(y)$, we can decide in $O(1)$ time whether $b$ precedes or succeeds $\beta$. 
The result of these two lemmas is an $O(\log n)$ algorithm to find $e^+(v)$ (Algorithm~\ref{alg:find-cqueue}). 

\begin{lemma}
\label{lem:decision}
    Let $v \in T$ have children $x$ and $y$. 
    Given edges $\alpha \in \mathbb{E}(x)$ and $\beta \in \mathbb{E}(y)$ with center points $l$ and $r$, respectively, we can decide whether $e^+(v)$'s endpoints succeed/precede $\alpha$ and $\beta$ in $O(1)$ time using only arithmetic and comparisons.
\end{lemma}

\begin{figure}[h]
\centering
\includegraphics[page=3]{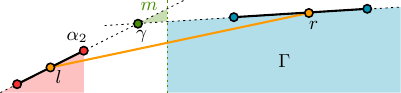}
\caption{
Lemma~\ref{lem:decision}, case 3.}
\label{fig:decisionproof3}
\end{figure}

\begin{proof}
The proof is a case distinction, depending on the slope of the segment $\overline{lr}$. 
Denote by $(a, b)$ the (unknown) endpoints of $e^+(v)$. We consider three cases (where the first two are not mutually exclusive) and their consequence:

\begin{figure}[t]
\centering
\includegraphics[]{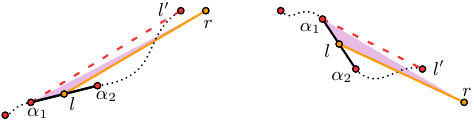}
\caption{
Lemma~\ref{lem:decision}, case 1, for upward and downward slopes.
The purple triangle denotes $\Delta$.
}
\label{fig:decisionproof}
\end{figure}

\textbf{Slope($\alpha$) $\leq$ Slope($\overline{lr}$): $a$ must precede $\alpha$.}
Denote by $(\alpha_1, \alpha_2)$ the endpoints of $\alpha$.
Consider the triangle $\Delta = (\alpha_1, l, r)$. Since slope($\alpha$) $\leq$ slope($\overline{lr})$, and $r$ per construction has a $x$-coordinate greater or equal to that of $\alpha_2$, the point $\alpha_2$ must be on or under the edge $\overline{lr}$ of this triangle (Figure~\ref{fig:decisionproof}). 
Per definition of convexity, the bridge $e^+(v)$ bounds a halfplane that contains all points in $CH^+(\pi(v))$. Thus, the bridge $e^+(v)$ bounds a halfplane that must contain $\Delta$.

We claim that any point $l'$ in $\pi(x)$ right of $\alpha_1$ cannot be an endpoint of $e^+(v)$.
Indeed, any such point $l'$ must lie on or above the line through $\overline{\alpha_1 r}$ (since the halfplane bounded by $e^+(v)$ must contain both $a_1$ and $r$). 
This implies that we have found a $l' \in \pi(x)$ such that the segment $\overline{\alpha_1 l'}$ has a greater slope than $\alpha$: which contradicts the assumption that $\alpha$ was part of the upper convex hull of $\pi(x)$.

\textbf{Slope($\overline{lr}$) $\leq$ Slope($\beta$): $b$ must succeed $\beta$.}
This proof is symmetric to the previous case.

\textbf{Slope($\alpha$) $>$ Slope($\overline{lr}$) $>$ Slope$(\beta)$.}
Denote by $m$ a vertical line separating $\pi(x)$ and $\pi(y)$ (we can compute such an $m$ in constant time). 
Denote by $\gamma$ the intersection point between the supporting lines of $\alpha$ and $\beta$. We prove that:
\begin{itemize}[noitemsep]
    \item if $\gamma$ lies on or left of $m$, $a$ must succeed $\alpha$, and
    \item if $\gamma$ lies on or right of $m$, $b$ must precede $\beta$.
\end{itemize}
We prove the first subcase (Figure~\ref{fig:decisionproof3}), as the second subcase is symmetric.
Denote by $\Gamma$ the area that can contain $b$, the area bounded by the halfplane supporting $\beta$ and right of $m$.  
Because $\gamma$ lies left of $m$, $\Gamma$ is entirely contained in the halfplane bounded by the supporting line of $\beta$. 
Consider any vertex $l' \in \pi(x)$ preceding $\alpha$, the segment $\overline{l'b}$ must be contained in the halfplane bounded by $\alpha$. 
However, this implies that the segment $\overline{l'b}$ bounds a halfplane that excludes $\alpha_2$. Thus, $a$ must succeed $\alpha$. 
Note that if $a$ succeeds $\alpha$ (and, $a$ lies in the green area of the figure) then $b$ may succeed or precede $\beta$.
\end{proof}

\begin{lemma}
\label{lem:leaf}
        Let $v \in T$ be an internal node with children $x$ and $y$. 
        Given the left endpoint $a$ of $e^+(v)$ and any edge $\beta \in \mathbb{E}(y)$, we can decide whether the right endpoint $b$ of $e^+(v)$ succeeds or precedes $\beta$ in $O(1)$ time using only arithmetic and comparisons. 
\end{lemma}

\begin{proof}
    Denote by $r$ the center point of $\beta$. 
    The sequence of edges on the upper convex hull of $\pi(v)$ per definition has decreasing slope. 
    It immediately follows that if Slope($\overline{a r}$) $>$ Slope($\beta$) then $b$ must precede $\beta$. 
    Similarly, if Slope($\overline{a r}$) $<$ Slope($\beta$) then $b$ must succeed $\beta$.
    Finally, if Slope($\overline{a r}$) $=$ Slope($\beta$) then the points $a, r$ and the endpoints of $\beta$ are collinear.
    Since our bridge is the maximal segment bounding a halfplane that contains $\pi(v)$, $b$ must succeed $\beta$. 
\end{proof}

\begin{corollary}
\label{cor:find-cqueue}
Let $v \in T$ have children $x$ and $y$. 
Given $\mathbb{E}(x)$ and $\mathbb{E}(y)$ as balanced trees, we may compute the bridge $e^+(v)$ in $O(\log n)$ time (Algorithm~\ref{alg:find-cqueue}). 
\end{corollary}

 \begin{algorithm}[h]
    \caption{FindBridge(node $v$, $\alpha \in \mathbb{E}(x)$, $\beta \in \mathbb{E}(y)$)}
    \label{alg:find-cqueue}
    \begin{algorithmic}[1]
      \WHILE{ ! ($\alpha$.isLeaf \AND $\beta$.isLeaf)}
      \STATE $(l, r)$ $\gets$ $(\alpha, \beta)$.getCenter()
      \IF{ Slope($\alpha$) $\leq$ Slope($\overline{lr}$) \AND !$\alpha$.isLeaf}
      \STATE $\alpha$ $\gets$ $\alpha$.leftChild
      \ENDIF
      \IF{ Slope($\overline{lr}$) $\leq$ Slope($\beta$) \AND !$\beta$.isLeaf}
      \STATE $\beta$ $\gets$ $\beta$.rightChild
            \ENDIF
            \IF{  neither first two cases apply}
            \IF {$\beta$.isLeaf}
            \STATE $\alpha$ $\gets$ $\alpha$.rightChild
            \ELSIF{$\alpha$.isLeaf}
            \STATE $\beta$ $\gets$ $\beta$.leftChild
            \ELSE
            \STATE $\gamma$ $\gets$ intersectLines($\alpha$, $\beta$)
            \STATE $m$ $\gets$ verticalLineSeparating$(\pi(x), \pi(y))$
            \IF{ $\gamma$ left of $m$}
            \STATE $\alpha$ $\gets$ $\alpha$.rightChild
            \ELSE
            \STATE $\beta$ $\gets$ $\beta$.leftChild
            \ENDIF            
            \ENDIF
            \ENDIF
      \ENDWHILE
      \RETURN ($\alpha$, $\beta$)
    \end{algorithmic}
  \end{algorithm}

\subsection{The update algorithm using $c$-queues}
\label{sec:cqueue} 
Now, we are ready to present our implementation of the dynamic algorithm in~\cite{overmars1980dynamically}, using our edge-based logic. We call this algorithm OvL. 
 Let us insert or delete a point $p \in P$ and denote by $\rho$ the root-to-leaf path to $p$ in $T$. 
 We compute for all $v \in \rho$ the bridge $e^+(v)$. 
This allows us to restore all c-queues, and store $CH^+(P)$ in $\mathbb{E}^*(r)$ (where $r$ is the root of $T$). 
Our algorithm works in two steps, that both take $O(\log^2 n)$ time:

\paragraph{Traversing down.}
We start at the root $r$ where $\mathbb{E}^*(r) = \mathbb{E}(r)$. 
We traverse $\rho$ from the root to $p$. 
For each node $v$, we assume we have $\mathbb{E}(v)$ as a balanced binary tree.
Let $p$ lie in the left child $x$ of $v$ ($p \in \pi(x)$). 
We invoke splitHull($x$, $\mathbb{E}(v)$): destroying $\mathbb{E}(v)$ and storing $\mathbb{E}_R(v)$ in $v$. 
The splitHull function gives us $\mathbb{E}(x)$ in $O(\log n)$ time, and with $\mathbb{E}(x)$ we  recurse on $x$.
\paragraph{Bubbling up.}
For the leaf $l$ that contains $p$, the bridge $e^+(l)$ equals $(p, p)$. 
We create or delete this bridge accordingly. 
From the downwards traversal in the previous step, we have the tree $\mathbb{E}(l)$ at $l$. 
We traverse $\rho$ from $l$ to the root. 
For each node $v$, we compute $\mathbb{E}(v)$ as follows. Let $x$ and $y$ be the children of $v$. If $x$ is the child on the path, we  computed $\mathbb{E}(x)$ before moving up to $v$ and we have $\mathbb{E}(y)$ from our downwards traversal. We first invoke Algorithm~\ref{alg:find-cqueue}  
to compute $e^+(v)$ in $O(\log n)$ time (Corollary~\ref{cor:find-cqueue}). We then compute $\mathbb{E}(v)$ joining $\mathbb{E}(x)$ and $\mathbb{E}(y)$ around the bridge $e^+(v)$.
 



\begin{theorem}
    Let $P$ be a  two-dimensional point set.  
    We can maintain the edges of $CH^+(P)$ subject to insertions and deletions in $P$ as a balanced binary tree $\mathbb{E}(r)$ in $O(\log^2 n)$ worst-case time per update. 
\end{theorem}

The balanced binary tree $\mathbb{E}^*(r)$ at the root stores the $h$ edges of $CH^+(P)$.
Queries~\ref{it:extreme}--\ref{it:hullintersect} can subsequently be answered in $O(\log h)$ time using the standard search algorithms over $\mathbb{E}^*(r)$~\cite{chan2011three}.

\section{Improving the Solution}
The update algorithm presented in \cite{overmars1980dynamically} differs from our new approach in the following choices:
\begin{itemize}[noitemsep]
    \item It traverses the path $\rho$ in $T$ twice: once top-down to ensure that we can compute for all $v \in \rho$ the tree $\mathbb{E}(v)$, and once bottom-up to compute the bridges. 
    \item It stores points in $P$ multiple times (both in $T$ and in the c-queues) requiring either double the space or pointers that point to non-contiguous data. 
    \item It uses the split and join operations on binary trees. Although these have $O(\log n)$ theoretical running time, they are inefficient in practice.
\end{itemize}

In Appendix~\ref{sec:without}, we resolve these issues. 
Our algorithm no longer maintains the convex hull $CH^+(P)$ in a balanced binary tree. 
Instead, we maintain a new structure that we call the Partial Bridge Tree (PBT). 
\begin{definition}
\label{def:PBT}
    The \emph{Partial Bridge Tree (PBT)} stores a point set $P$ in a leaf-based balanced binary tree $T$ (sorted by $x$-coordinate).
Each node $v \in T$  stores  $e^+(v)$ and $e^-(v)$.
\end{definition}

\noindent
We can cleverly navigate our Partial Bridge Tree to get two properties:
\begin{enumerate}[noitemsep]
    \item $\forall v \in T$, given an edge $e \in CH^+( \pi(v) )$, we can find its successor on $CH^+( \pi(v))$ in $O(\log n)$ time. \label{it:prop1}
    \item $\forall v \in T$, given an edge $e \in CH^+( \pi(v) )$, we can find the median edge of all edges preceding $e$ on  $CH^+(\pi(v))$, in $O(\log n)$ time. \label{it:prop2}
\end{enumerate} 

The first property allows us to report $CH^+(P)$ in $O(h \log n)$ time. 
The second property naively allows us to answer Queries~\ref{it:extreme}--\ref{it:hullintersect} in $O(\log^2 n)$ time. 
Hence, we obtain an output that is somewhere in between the implicit and explicit convex hull.

\paragraph{Maintaining the Partial Bridge Tree.}
The second property naively also gives us an $O(\log^3 n)$ update time to maintain the Partial Bridge Tree. 
Indeed, consider after inserting or deleting point $p$ in $P$ the path $\rho$ from $p$ to the root of $T$. 
To restore the Partial Bridge Tree, we need to compute for all $v \in \rho$ (with children $x$ and $y$) the bridge $e^+(v)$. 
Algorithm~\ref{alg:find-cqueue} requires as input $v$ plus two edges $(\alpha, \beta) \in \mathbb{E}(x) \times \mathbb{E}(y)$. 
We then recurse by replacing $\alpha$ with $\alpha'$ (the median successor, or predecessor, on $\mathbb{E}(x)$). 
In Section~\ref{sec:cqueue}, we obtained $\alpha'$ in $O(1)$ time because we had $\mathbb{E}(x)$ as a balanced binary tree. 
Using Property~\ref{it:prop2}, we may instead get $\alpha'$ in $O(\log n)$ time.
Thus, we obtain an update algorithm with $O(\log^3 n)$ update time and $O(\log^2 n)$ query time just by replacing the functions `$\alpha$.leftChild' and `$\alpha$.rightChild' with the algorithm from Property~\ref{it:prop2}. 

We can, however, do better. We show that invoking Property~\ref{it:prop2} $O(\log n)$ times takes $O(\log n)$ worst-case total time. 
This creates a new algorithm for dynamic convex hull which we call \textbf{Eilice}. 
Eilice has a worst-case update time of $O(\log^2 n)$, matching OvL~\cite{overmars1980dynamically}, and a query time of $O(\log n)$. 
This update algorithm no longer requires us to maintain c-queues. 
In addition, we no longer require the initialization step of OvL: ensuring that we traverse $T$ exactly once. Thus, it circumvents the downsides of OvL at the cost of query time efficiency.

\section{Rank-based Convex Hulls}\label{sec:rankbased}



Finally, we consider rank-based convex hulls. 
Let $Y$ be a set of values, where the rank of $y \in Y$ is its index in the sorted order. 
We denote by $P_Y$ the two-dimensional point set that is obtained by mapping each value in $Y$ to $(\textsc{rank}, \textsc{value})$ and wish to dynamically maintain $CH(P_Y)$. 
The problem in this setting is that after inserting into or deleting from $Y$, the $x$-coordinate of linearly many points in $P_Y$ changes. 
Changing a value $y$ may change $CH(P_Y)$ by $\Theta(n)$ edges, even if $y$ itself was not on the convex hull (Figure~\ref{fig:hullchange}). 
The key observation to maintaining the convex hull in this setting is the following. 
After updating an element $y \in Y$ a  bridge $e^+(v)= (a, b)$ in the Partial Hull Tree is updated if and only if $y$ is in the subtree rooted at $v$. That is, if $y$ is not in the subtree rooted at $v$ then the $x$-coordinates of $(a, b)$ may both increase or decrease by one, but the bridge $e^+(v)$ remains a segment between the same two values.
Since the convex hull is implied by the set of all bridges in $T$, we may still maintain the Partial Hull Tree with the previous root-to-leaf update strategy. 

\paragraph{Implicit bridges}
In a Partial Hull Tree $T$, the leaves store the values in $P_Y$, sorted by $x$-coordinate. 
I.e., we store the values of $Y$ in the leaves of $T$ in their stored order.
For a node $v$ with children $x$ and $y$, the bridge $e^+(v)$ is the bridge between the convex hulls $CH^+(\pi(x))$ and $CH^+(\pi(y))$. 
For a bridge $e^+(v)$, we can no longer store the endpoints of the bridge explicitly: as the $x$-coordinate of all bridges may radically change after an update in $Y$. 
We define the implicit bridge $\eps^+(v)$ which stores only the two values $(y_1, y_2)$ corresponding to the endpoints of $e^+(v)$. 
At this point, we wish to note that we can easily maintain the Partial Hull Tree  using implicit bridges with a factor $O(\log n)$ overhead.
Indeed, we may run any of the two proposed algorithms. 
Whenever we need to consider a bridge $\alpha = e^+(x)$, we can get the corresponding values $(y_1, y_2)$ from the implicit bridge.
Then, we may perform a binary search over $Y$ to obtain their corresponding ranks.
Thus, at $O(\log n)$ overhead, we always have explicit access to the endpoints of $\alpha$.

In Appendix~\ref{app:rankedbased} we show that we can cleverly navigate $T$ to avoid this overhead. 
Our key contribution is that we show that we can perform a similar trick as for Eilice: performing $O(\log n)$ such rank queries in $O(\log n)$ total time. This gives the first dynamic algorithm for rank-based convex hulls with worst-case $O(\log^2 n)$ update time.
Our solution has $O(\log h)$ query time (used as an extension on OvL) or $O(\log n)$ query time (when applied as an extension of Eilice). 

\section{Experiments 
}
\label{sec:experiments}


In order to examine the empirical performance of our data structures, we perform a series of tests across several types of input, comparing the performance of our structure variants with each other, and to state of the art implementations of static convex hull algorithms.

\paragraph{Test environment and implementations.}

For experiments we consider the following implementations:
\begin{description}[noitemsep]
    \item[Simplified OvL] our simplified Partial Hull Tree structure using concatenable queues. 
    \item[Eilice] our simplified Partial Hull Tree structure without concatenable queues. 
    \item[CHHN] a Java implementation of the original Overmars and van Leeuwen structure~\cite{javaimpl}.
    \item[Eddy] CGAL implementation of Eddy's algorithm~\cite{impleddy} ($O(nh)$ construction time).
    \item[GA] CGAL implementation of Andrew's variant of the Graham scan algorithm~\cite{implga} ($O(n\log n)$ construction time).
\end{description}

The implementations of our data structures are available at~\cite{ourimpl}. As a robustness check, we have compared the points we have found on the hull to those found by static convex hull algorithms implemented in CGAL, and verified that they are in agreement.

We note that comparing implementations across programming languages is inherently a dubious task. Not only do the languages differ in their approach to things like memory management, which can greatly influence practical performance, but in this case, the comparisons are further troubled by the fact that geometric robustness has not been implemented for \emph{CHHN}. The comparisons between the Java and C++ implementations should therefore not be seen as direct performance comparisons, but rather the Java implementation acts as an indicator of existing solutions.

The experiments are run on a cluster node with an Intel Xeon Gold 6226R CPU on a single core at 2.8GHz.

\paragraph{Test data.} We generated four distinct data sets of $2^{20}$ points each, following Gamby and Katajainen~\cite{gamby2018convex} who perform experimental analysis on static convex hull algorithms. The first three categories are from~\cite{gamby2018convex}, and the last is tailored to measure behaviour when all points lie on the convex hull. Recall that $h$ denotes the number of points in $CH(P)$. We consider the following data sets:
\begin{description}[noitemsep]
    \item[Uniformly random data] each point of $P$ is drawn uniformly at random from a  square; thus, the expected number of extremal points is $\Theta(\log n)$. 
    \item[Disk-truncated data] where each point is drawn uniformly at random, but only added to $P$ when it lies in a pre-specified disk. Here, the expected number of extremal points is $\Theta(n^{1/3})$.
    \item[Bell data] where data is generated according to a normal distribution (and thus, we expect $\Theta(\sqrt{\log n})$ extremal points).
    \item[Circular data] where points are sampled from the boundary of a pre-specified disk; i.e. every data point belongs to the hull. 
\end{description}

For the uniformly random and Bell data points, we follow the specifications of \cite{gamby2018convex}. For the disk-truncated and circular points, we consider disks of radius $1000$ centered at $(0,0)$.
We note that in all cases, contrary to Gamby and Katajainen~\cite{gamby2018convex}, we do not restrict the points to have integer coordinates. This restriction served to avoid correctness issues that can arise in practice when comparing floating point values of finite precision. Integration with the CGAL geometry kernel allows us to circumvent these geometric robustness issues, at some cost of performance, opening up for use in applications that do not allow for such a restricted input.

\paragraph{Experiments.} The contribution of this work is an updatable, queriable convex hull implementation for two-dimensional and rank-based datasets. We thus test for the following categories:
\begin{description}[noitemsep]
    \item[Dynamic construction] a simulated dynamic scenario in which we construct the hull of a point set by repeated extension. 
    \item[Extension] the time it takes to extend the point set $P$ with an additional point set and obtain the convex hull of their union.
    \item[Queries] the time it takes to perform a sequence of point-containment queries on an existing hull.
    \item[Updates] the time it takes to perform an interspersed sequence of insertions and deletions on an existing hull.
\end{description}

The dynamic construction test serves to emulate behaviour one might encounter in a typical dynamic setting. The hull is repeatedly extended, so that queries can be answered often on the current state of the hull. This brings a clear disadvantage to the static algorithms, and thus we examine how various rates of extension affect their performance.

For extensions, the static algorithms simply rebuild the hull. This is to determine if the dynamic variations can compare to simply rebuilding the hull using a static algorithm, eliminating the need for a dynamic structure in cases where queries are needed on an irregular basis. Each measurement of extension adds $50,000$ new points, while each set of updates consists of $1000$ updates split evenly between insertions and deletions. There is no overlap between inserted and deleted points, and the order of updates is random, emulating the updates pattern one might encounter in practical applications.

Due to the speed at which queries can be answered, we for each measurement perform $2^{20}$ queries for points chosen according to a uniform distribution over a bounding box of the area in which the full test data resides.

\paragraph{Results and discussion.}

\begin{figure}[t]
\centering
\includegraphics[width=.99\linewidth]{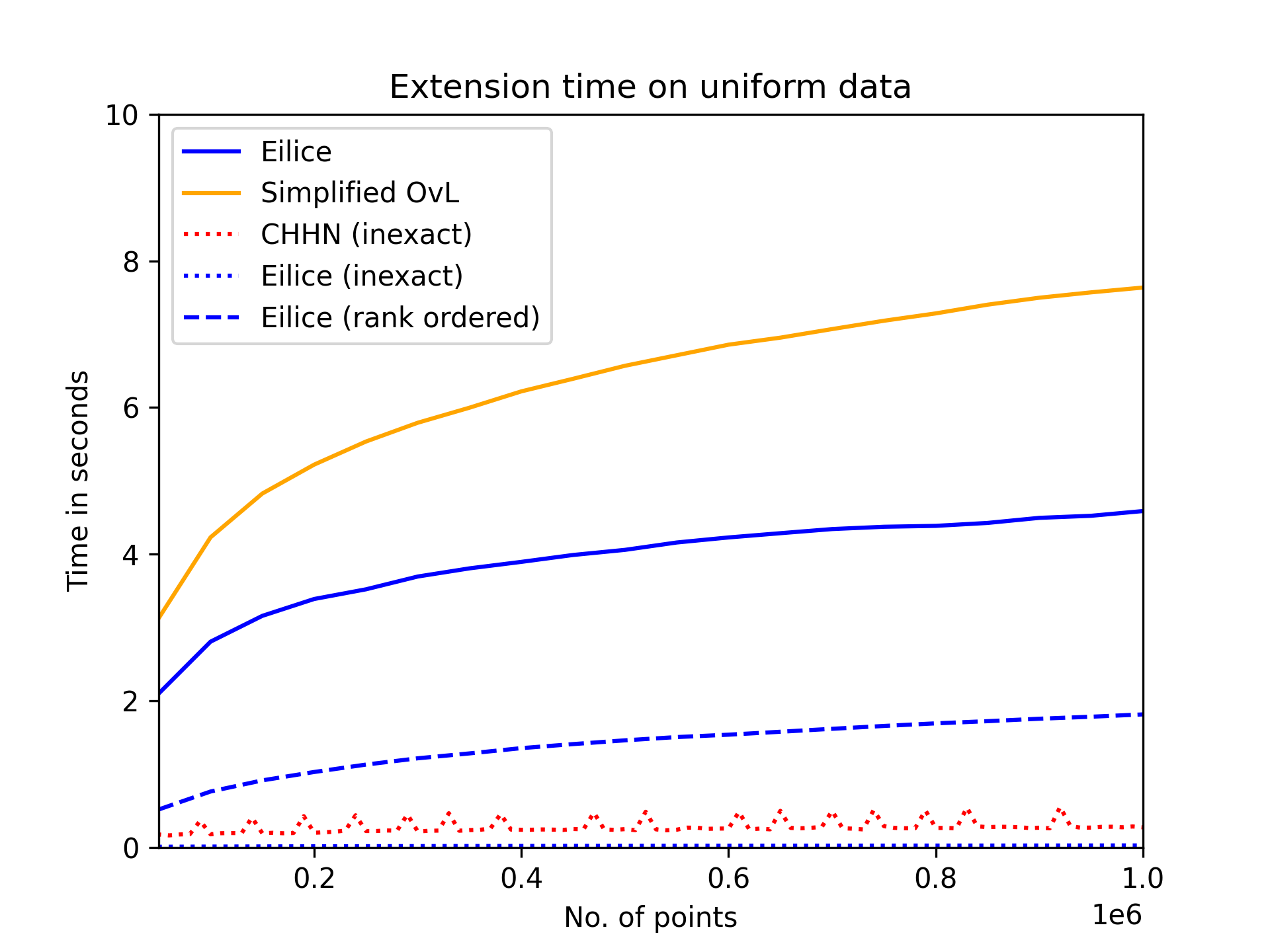}
\caption{
    Extension time on uniform data.
    }
\label{fig:extend_unif}
\end{figure}

For ease of readability, we highlight only tests on specific data sets. The remainder can be found in Appendix~\ref{app:results}, and the raw results can be found in~\cite{ourimpl}.

\begin{figure}[htb]
\centering
\includegraphics[width=.99\linewidth]{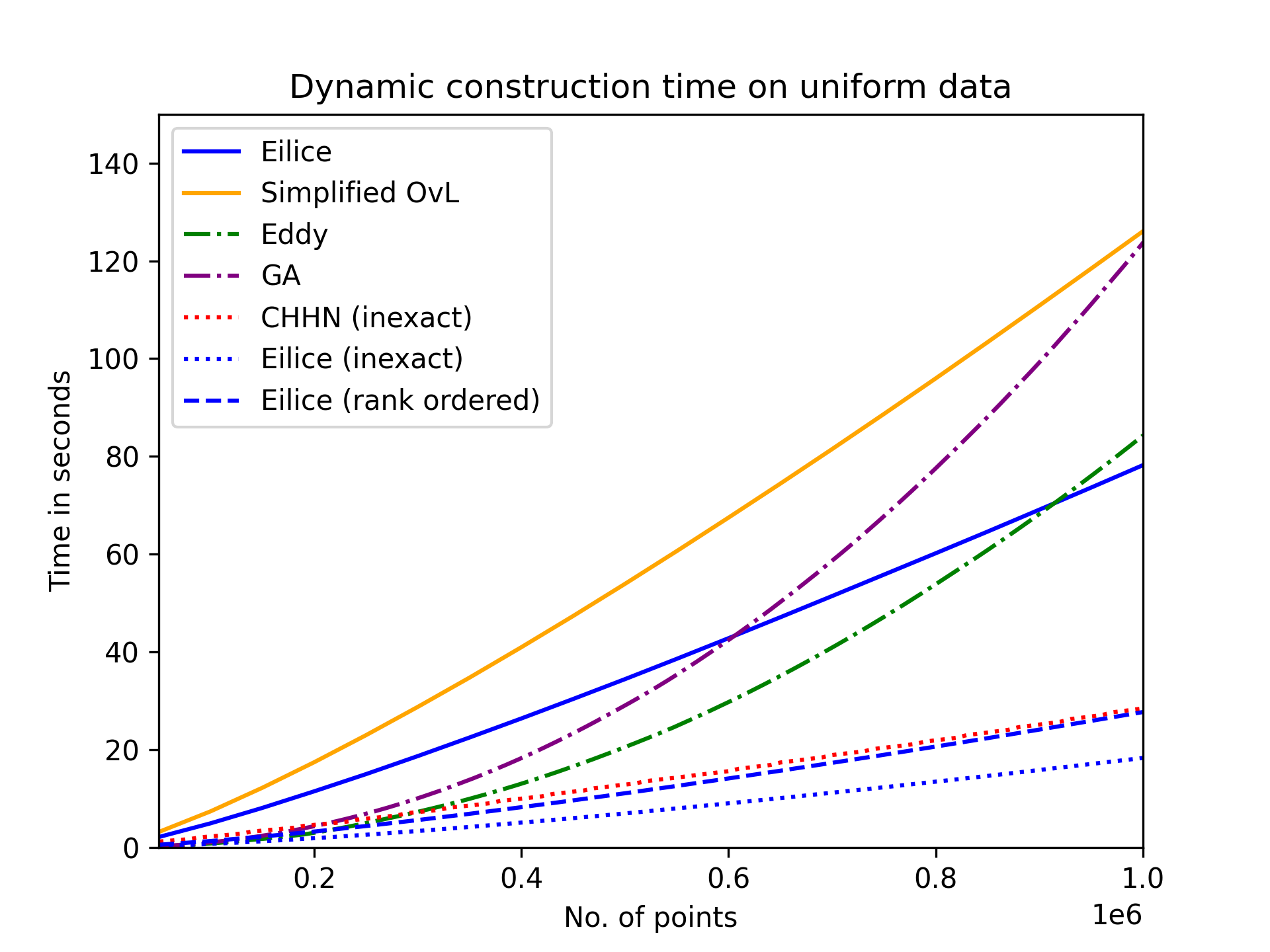}
\caption{
    Dynamic construction time using $500$ point extensions on uniform data.\vspace{-0.5cm}
    }
\label{fig:construct_unif}
\end{figure}

In Figure~\ref{fig:construct_unif} we show the dynamic construction test using $500$ point extensions. At roughly a million points and forward we see that exact Eilice beats the static variations, while the inexact implementations take over sooner.

\begin{figure}[htb]
\centering
\includegraphics[width=.99\linewidth]{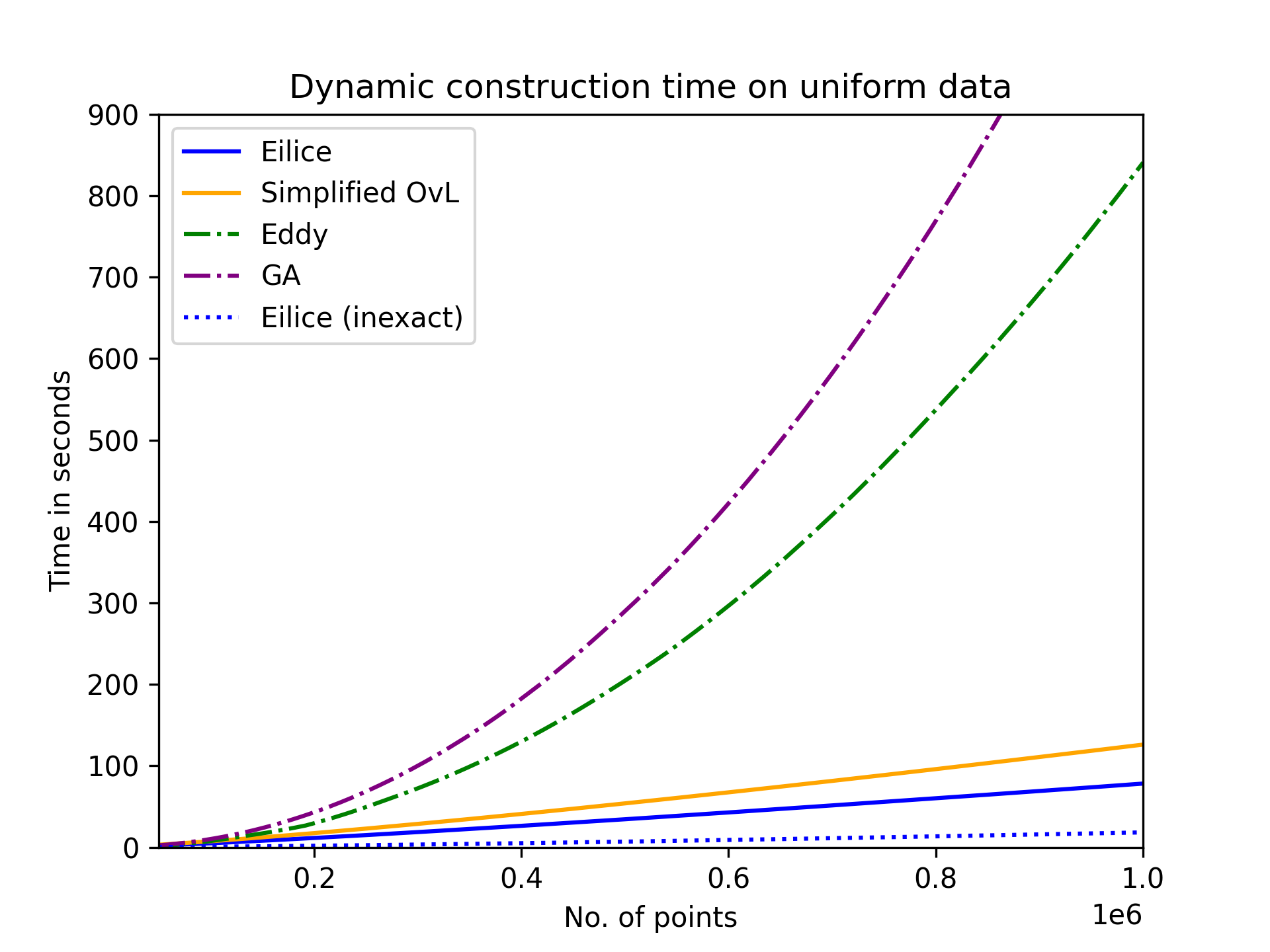}
\caption{
    Dynamic construction time using $50$ point extensions on uniform data.\vspace{-0.5cm}
    }
\label{fig:construct_unif_50}
\end{figure}

In practical applications, one's access pattern might require access to the hull more often than once every $500$ updates. For a scenario with access every $50$ updates we get the results seen in Figure~\ref{fig:construct_unif_50} where the clear benefit of a dynamic structure can be seen.

To explicitly showcase the difference in performance when comparing exact and inexact implementations, we show in Figure~\ref{fig:extend_unif} the extension time on uniform data. What should be noted here is the gap between the inexact and exact data structures. Replacing simple floating point comparisons with more robust and involved operations takes a large toll on performance. This is perhaps not surprising, but it is necessary if one wants to guarantee correct comparisons, and thus correct behaviour. We also show the time spent extending the rank-ordered implementation. This is because rank-ordered integers avoid some of the pitfalls of geometric robustness, and thus fall into a similar category as the inexact implementations. Of note is that the rank-ordered implementation has to do additional bookkeeping, placing it somewhat in the middle of the extremes. We also note the logarithmic nature of the exact dynamic implementations.
The jagged curve of CHHN does not stem from the usage of an amortized structure, but rather the Java language's built-in garbage collector periodically performing memory cleaning.
The static algorithms have a large advantage here, requiring a single reconstruction for every $50,000$ insertions, placing them far ahead of most of the competition. 

\begin{figure}[t]
\centering
\includegraphics[width=.99\linewidth]{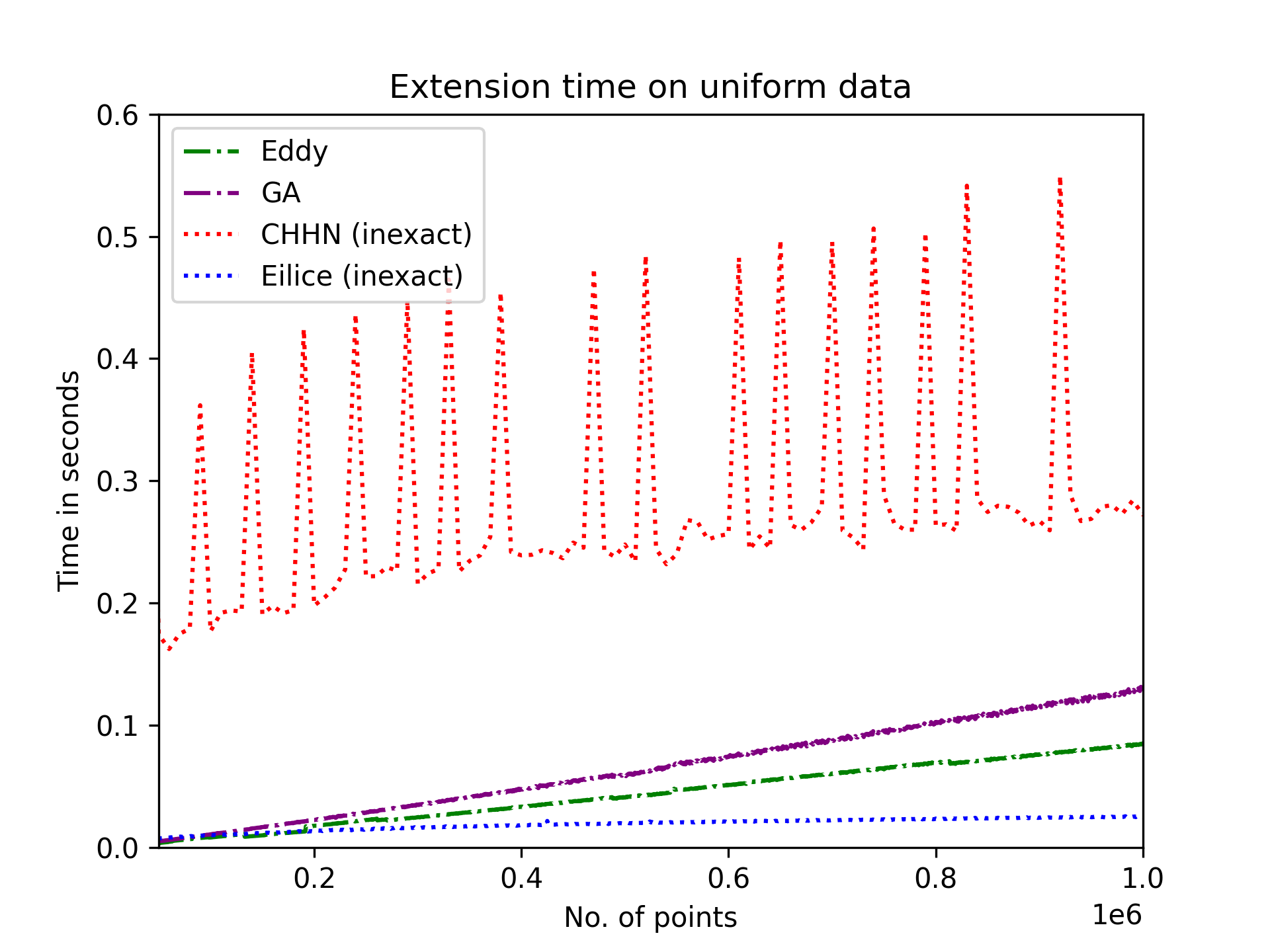}
\caption{
    Extension time on uniform data of select implementations.\vspace{-0.5cm}
    }
\label{fig:extend_inexact}
\end{figure}

To clarify the behaviour of extension we also show in~Figure~\ref{fig:extend_inexact} the behaviour excluding the exact dynamic implementations. Although tempting to simply claim that our simplification causes a speedup of an order of magnitude compared to the regular Overmars and van Leuween structure, the difference here between Java and C++ muddles any conclusions.
Here it is also clearer that the static algorithms have super-logarithmic growth, however, the constants associated with the exact implementations mean that data has to grow much larger for the exact dynamic implementations to extend faster.

\begin{figure}[t]
\centering
\includegraphics[width=\linewidth]{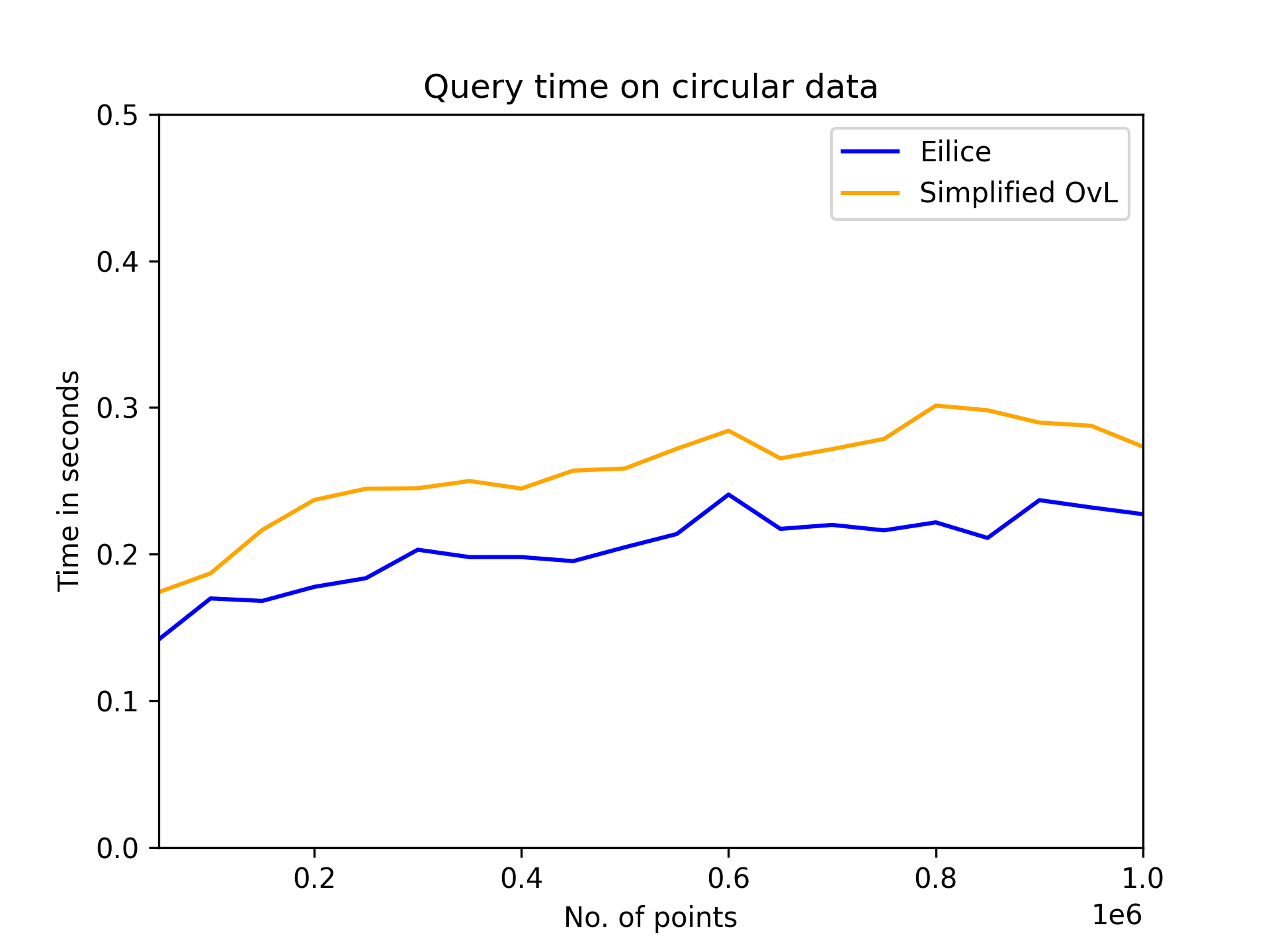}
\includegraphics[width=\linewidth]{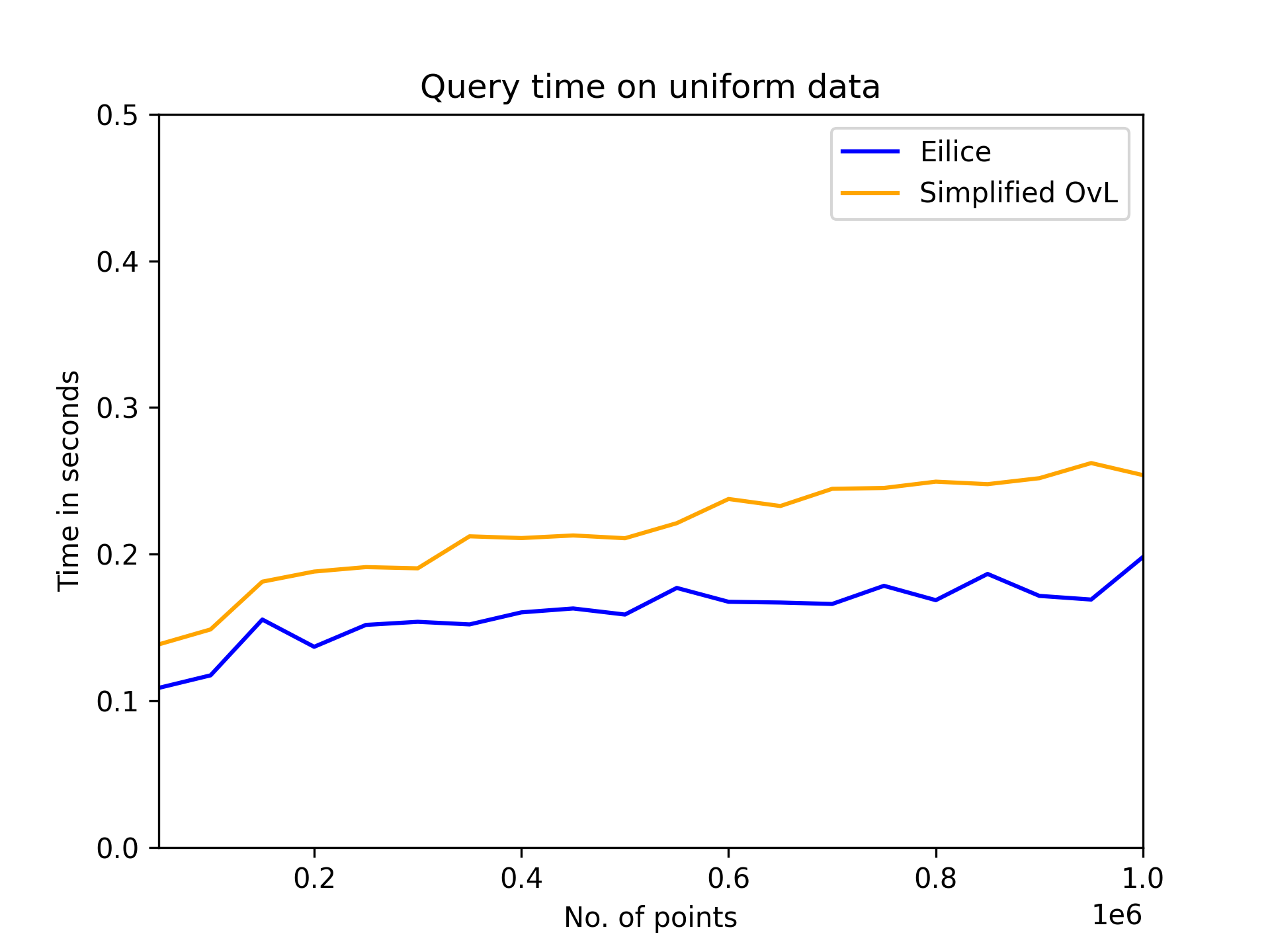}
\caption{
    Query time on circular and uniform data.
    }
\label{fig:query}
\end{figure}

For the queries, we specifically consider the circular and uniform random data. Recall that the expected number of points on the hull of the uniform data, $h$, is $O(\log n)$ for this point set. 
The benefit of maintaining concatenable queues is that they provide the ability to answer queries in $O(\log h)$ time rather than worst-case $O(\log n)$. In Figure~\ref{fig:query}, however, we see that the Eilice algorithm obtains similar performance to the simplified OvL solution, even whilst not having concatenable queues. 
This behaviour can be explained as follows: 
the Eilice query algorithm considers an edge $e^+(v) = \alpha$ on the upper convex hull.
Then, Eilice either outputs the correct answer or considers the left (or right) child $x$ of $v$.
If the Eilice algorithm does not output the correct answer, there exist two cases. Either the bridge $e^+(x)$ is also on the convex hull, or the $x$-coordinate of $e^+(x)$ succeeds the left endpoint of $\alpha$. The Eilice algorithm makes a comparison that the simplified OvL does not make, \emph{only} in the latter case. 
Due to how the tree is defined and constructed, that case is rare. Thus, in practice, the Eilice and simplified OvL algorithms oftentimes execute the same decision tree before reaching the desired output (even though Eilice has a worse worst-case guarantee).

\begin{figure}[t]
\centering
\includegraphics[width=.99\linewidth]{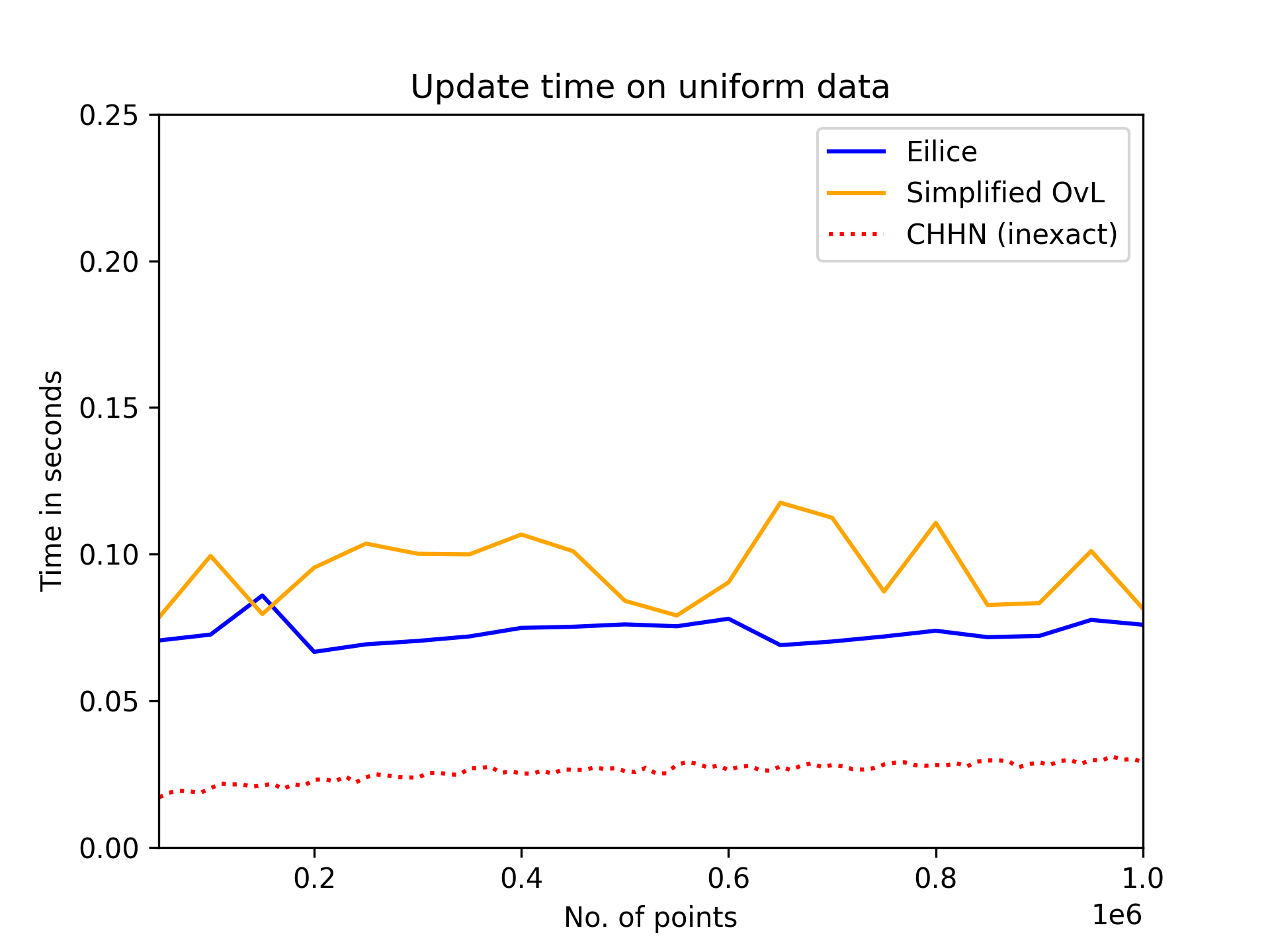}
\caption{
    Time spent on an interspersed sequence of $500$ insertions and $500$ deletions on an existing convex hull on uniform data.
}
\label{fig:update}
\end{figure}

For the update measurements, shown in Figure~\ref{fig:update}, we again see the logarithmic behaviour for both dynamic algorithms. As expected, Eilice beats simplified OvL by virtue of not having the overhead associated with maintaining concatenable queues.

Our experiments highlight several key points when deciding what algorithms to employ for dynamic convex hulls in practice. If geometric robustness is not necessary, one can achieve much greater performance. This might be the case in applications that accept erroneous queries, in which data is inherently already unreliable, or where data happens to be distributed enough that errors become unlikely. As an example of the latter, the uniform random data set was correctly handled by the inexact data structures, while the circular data was much more prone to precision errors.

If geometric robustness is necessary, but queries can be relegated to infrequent batches, static recomputations might be preferred, if the queries are infrequent enough or the data sizes are large enough.

If both geometric robustness and dynamic behaviour is required, or data is rank-based, Eilice is the clear winner. While leaving out the concatenable queues gives worse worst-case guarantees, it requires an adversarial query pattern for the overhead of Eilice to become worse.

\paragraph{To conclude:} 
We simplify the bridge-finding procedure of Overmars and van Leuween, by a clever change from vertex based to edge-based computations. This simplification leads to much less error-prone implementations, at no practical cost. 
We implement and test our simplified data structure, along with an additional trade-off between queries and updates, by leaving out concatenable queues. The implementations are publicly available at~\cite{ourimpl}.
We examine the practical implications of these simplifications, providing insight into not only how our structure performs, but also what factors one should consider when employing a dynamic convex hull data structure, with guidance for various use cases.

Finally, the simplified algorithm also easily extends to rank-ordered data, accommodating updates and queries in polylogarithmic time, thus giving the first data structure for dynamic maintenance of convex hulls of rank-ordered data with non-trivial update bounds.

\newpage


\bibliographystyle{plain}
\bibliography{refs}

\appendix

\newpage
\section{Eilice: updates without c-queues}
\label{sec:without}

The update algorithm OvL has three components that make it practically less efficient:
\begin{itemize}[noitemsep]
    \item It traverses the path $\rho$ in $T$ twice: once top-down to ensure that we can compute for all $v \in \rho$ the tree $\mathbb{E}(v)$, and once bottom-up to compute the bridges. 
    \item It stores points in $P$ multiple times (both in $T$ and in the c-queues) requiring either double the space or pointers that point to non-contiguous data. 
    \item It uses the split and join operations on binary trees. Although these have $O(\log n)$ theoretical running time, are very inefficient in practice.
\end{itemize}

In this subsection, we show that we may avoid maintaining c-queues and these  downsides by navigating only $T$. As a result, we obtain different outputs than the OvL algorithm. 
We no longer are able to store $CH(P)$ in a balanced binary tree. 
Instead, we ensure that at all times we can report the $h$ points on $CH(P)$ in $O(h \log n)$ time. 
In addition, we show that we facilitate convex hull queries in $O(\log n)$ time as opposed to $O(\log h)$ time.

\paragraph{Key definitions.}
Definition~\ref{def:PBT} gives the definition of our data structure: the Partial Bridge Tree $T$ (PBT).
Let $v \in T$ be an internal node, then per definition, $v$ has two children. 
The \emph{median} $med(v)$ is the leftmost point in the subtree of the right child of $v$ (see Figure~\ref{fig:fourcases} (a) ). Our key observation is that for every node $v$: the bridge $e^+(v)$ (when projected onto the $x$-axis) is an $x$-interval that contains $med(v)$:

\begin{figure}[t]
\centering
\includegraphics[]{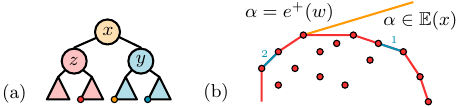}
\caption{
(a) A node $v$ with children $x$ and $y$. We show the medians as vertices with the same colour as the corresponding node.  
}
\label{fig:fourcases}
\end{figure}

\begin{lemma}
\label{lem:step_size}
Let $v$ be a node in $T$ and $e$ be the corresponding bridge. 
Then $e^+(v)$ either has $med(v)$ as its right endpoint or has two endpoints that precede and succeed $med(v)$, respectively.
\end{lemma}

\begin{proof}
Let $v$ have children $x$ and $y$. 
The lemma follows from the fact that $med(v)$ is the leftmost point in $\pi(y)$, all points in $\pi(x)$ precede $med(v)$,  and $e^+(v) = (a, b) \in \pi(x) \times \pi(y)$. 
\end{proof}

We may use this lemma, together with the following observation to obtain an equivalent to Corollary~\ref{cor:find-cqueue}:

\begin{observation}
\label{obs:leftchild}
Let $w \in T$ and denote by $\alpha = e^+(w) = (a, b)$.
For any vertex $l$ in the left subtree under $w$ with bridge $\gamma = e^+(l)$, 
there are two cases (see Figure~\ref{fig:fourcases}(b)):
\begin{enumerate}[noitemsep]
    \item The left endpoint of $\gamma$ succeeds $a$.
    Then all bridges in the right subtree of $l$ are either not on $\mathbb{E}(x)$ (as they are `covered' by $\alpha$) or also succeed $a$.
    \item The right endpoint of $\gamma$ precedes $a$. Then $\gamma \in \mathbb{E}(x)$.   
\end{enumerate}
\end{observation}

\begin{lemma}
\label{lem:no-cqueue}
    Let $v \in T$ have children $x$ and $y$. 
Recall that $\mathbb{E}(x)$ and $\mathbb{E}(y)$ are the convex hull edges of $CH^+(\pi(x))$ and $CH^+(\pi(y))$ in their cyclical ordering. 
Given edges $\alpha = e^+(x)$ and $\beta = e^+(y)$, we may compute the bridge $e^+(v)$ in $O(\log n)$ time (Algorithm~\ref{alg:find-cqueue}+\ref{alg:leftChild}). 
\end{lemma}

\begin{proof}
    We no longer have access to $\mathbb{E}(x)$ and $\mathbb{E}(y)$ as balanced binary trees. 
    In Algorithm~\ref{alg:find-cqueue}, we consider edges $(\alpha, \beta) \in \mathbb{E}(x) 
    \times \mathbb{E}(y)$ and we use Lemma~\ref{lem:decision} to decide whether the bridge endpoints $e^+(v)$ precede or succeed $\alpha$. 
    Suppose that the left endpoint of $e^+(v)$ precedes $\alpha$ (all other scenarios are symmetrical). 
    Then, we replace $\alpha$ by an edge $\alpha' \in \mathbb{E}(x)$ that precedes $\alpha$. In Section~\ref{sec:with-cqueue}, we obtain $\alpha'$ by simply replacing $\alpha$ with its left child in the balanced binary tree on $\mathbb{E}(x)$.

    Since we no longer have this binary tree, for any edge $\alpha \in \mathbb{E}(x)$ its left child (or right child) is no longer well-defined. Thus all code lines that say $\alpha$.leftChild (or $\alpha$.rightChild) were no longer well-defined. 
    However, for all $\alpha \in \mathbb{E}(x)$ there exists a node $w \in T$ where $\alpha = e^+(w)$. We now note that Algorithm~\ref{alg:leftChild} gives us an edge $\alpha'$ preceding $\alpha$ on $\mathbb{E}(x)$. 

    This leads to a simple alteration of Algorithm~\ref{alg:find-cqueue}.
    Note that the function $\alpha$.rightChild() can be defined analogously. 
    We replace all calls:
    \begin{itemize}[noitemsep]
        \item $\alpha$.leftChild by $\alpha$.leftChild(),
        \item  $\beta$.leftChild by $\beta$.leftChild(),
        \item $\alpha$.rightChild by $\alpha$.rightChild(),
        \item  $\beta$.rightChild by $\beta$.rightChild().
    \end{itemize} 

    In each iteration of the while loop, either $\alpha$ is a leaf in $T$ or an edge on $\mathbb{E}(x)$.
    We may call invoke the function leftChild() at most $O(\log n)$ times until we reach a leaf of $T$.
    Thus, Algorithm~\ref{alg:find-cqueue} terminates in $O(\log n)$ time and computes the bridge $e^+(v)$. 
\end{proof}

 \begin{algorithm}[h]
    \caption{$\alpha$.leftChild() for $\alpha = e^+(w)$ and $\alpha \in \mathbb{E}(x)$}
    \label{alg:leftChild}
    \begin{algorithmic}[1]
        \STATE $l \gets $ the left child of $w$ in $T$ 
        \STATE $\gamma \gets e^+(l)$
        \IF{$\gamma$.x succeeds $\alpha$.x}
        \RETURN $l$.leftChild()
      \ENDIF
      \RETURN ($e^+(l)$, $l$)
    \end{algorithmic}
  \end{algorithm}

\newpage

\begin{lemma}
    Given the Partial Bridge Tree $T$, we can report the $h$ edges on $CH^+(P)$ in $O(h \log n)$ time. 
\end{lemma}

\begin{proof}
    This lemma follows almost immediately from Observation~\ref{obs:leftchild} and the leftChild() and rightChild() functions.
    Indeed, let $r$ be the root of $T$, then $e^+(r)  = \alpha \in \mathbb{E}(r) = CH^+(P)$. 
    We recursively invoke $e^+(r).leftChild()$ to obtain an edge $\beta \in  CH^+(P)$.
    All edges of $CH^+(P)$ in between $\alpha$ and $\beta= e^+(w)$, per definition succeed $\alpha$ and precede $\beta$. 
    Thus, we may recurse into the right subtree of $w$ to obtain vertices $x \in T$. 
    We may again apply Observation~\ref{obs:leftchild} to note that if $e^+(x) \in CH^+(P)$ if and only if it precedes $\beta$ and succeeds $\alpha$.
    This way, each time we explore a new subtree in $T$ we find at least one edge of $CH^+(P)$.
    Since our function recurses at most $O(\log n)$ times before it reaches a leaf of $T$ and thus we output $CH^+(P)$ in $O(h \log n)$ time. 
\end{proof}

\paragraph{Update algorithm.}
Let $p$ be a point that is inserted or deleted in $P$. 
We restore the Partial Bridge Tree as follows: we do a root-to-leaf traversal in $T$ to the leaf $l$ that contains $p$: spending $O(1)$ time per node.
For deletions, we may avoid this by following a pointer to the leaf in $T$.
Denote by $\rho$ the path from $l$ to the root of $T$. 
For all $v \in \rho$, bottom-up, we use Lemma~\ref{lem:no-cqueue} to compute $e^+(v)$ and this restores the Partial Bridge Tree. 
Thus, we have an $O(\log^2 n)$ update algorithm to maintain our Partial Bridge Tree.

\subsection{Supporting queries}
What remains is to show that our data structure can answer Queries (1)-(6) in $O(\log n)$ time.
The standard algorithms to answer these queries assume access to a balanced binary tree $\mathbb{E}(r)$ that stores the upper convex hull edges in their sorted order. 
Each of these algorithms subsequently does the following: they consider an edge $\alpha \in \mathbb{E}(r)$. They can either answer the query immediately using $\alpha$ or discard all edges of the convex hull preceding or succeeding $\alpha$. 
Doing this for both the upper and lower convex hull allows queries can be answered in $O(\log h)$ time where $h$ is the number of edges in $\mathbb{E}(r)$.
Using Algorithm~\ref{alg:leftChild} to navigate $T$ instead of $\mathbb{E}(r)$ allows our approach to immediately answer queries in $O(\log n)$ time instead. 
To illustrate this, we show how to answer Query $(4)$: deciding whether a query point $q$ lies in $CH(P)$.

\begin{lemma}
    Let $q$ be an arbitrary query point in $\mathbb{R}^2$. Using our data structure we can test if $q \in CH(P)$ using our Partial Hull Tree in $O(\log n)$ time.
\end{lemma}

\begin{proof}
Note that $q$ lies in $CH(P)$ if and only if it lies in both $CH^+(P)$ and $CH^-(P)$. 
    The point $q$ lies in $CH^+(P)$ if and only if there exists a unique edge $\gamma^* \in \mathbb{E}(r)$ where: 
    $q$ succeeds the left endpoint of $\gamma^*$, $q$ precedes the right endpoint of $\gamma^*$. Moreover, $q$ must lie below the line through $\gamma^*$.     
    Denote by $r$ the root of the Partial Hull Tree.
    
    Per definition, $e^+(r) \in \mathbb{E}(r)$. 
    Set $\alpha \gets e^+(r)$
    We test in $O(1)$ time whether $q$ succeeds the left endpoint of $\alpha$ and precedes the right endpoint of $\alpha$,
    If both conditions are true then $\gamma^* = \alpha$, we output whether $q$ is below the line through $\alpha$. 
    If $q$ precedes the left endpoint of $\alpha$ then $\gamma^*$ must precede $\alpha$ on the upper convex hull.
    Thus, we may step left from $\alpha$ using Algorithm~\ref{alg:leftChild}. 
    It follows that in $O(\log n)$ steps, we find $\gamma$ and answer the query accordingly.
\end{proof}

Maintaining our data structure, with the above query algorithms, implies the following theorem:

\begin{theorem}
    Let $P$ be a  two-dimensional point set.
    We can store $P$ in an $O(n)$ size data structure with $O(\log^2 n)$ worst-case time per update such that we may report the $h$ edges on $CH(P)$ in $O(h \log n)$ worst-case time.
    Moreover, we support all convex hull queries in $O(\log n)$ worst-case time.
\end{theorem}

\section{Ranked-based convex hulls}
\label{app:rankedbased}

In this section, we consider rank-based convex hulls. 
Let $Y$ be a set of values, where the rank of $y \in Y$ is its index in the sorted order. 
We denote by $P_Y$ the two-dimensional point set that is obtained by mapping each value in $Y$ to $(\textsc{rank}, \textsc{value})$ and wish to dynamically maintain $CH(P_Y)$. 
The problem in this setting is that after inserting into or deleting from $Y$, the $x$-coordinate of $O(n)$ points in $P_Y$ changes. 
Changing a value $y$ may change $CH(P_Y)$ by $O(n)$ edges, even if $y$ itself was not on the convex hull (Figure~\ref{fig:hullchange}). 
The key observation to maintaining the convex hull in this setting is the following. 
After updating an element $y \in Y$ a  bridge $e^+(v)= (a, b)$ in the Partial Hull Tree is updated if and only if $y$ is in the subtree rooted at $v$. That is, if $y$ is not in the subtree rooted at $v$ then the $x$-coordinates of $(a, b)$ may both increase or decrease by one, but the bridge $e^+(v)$ remains a segment between the same two values.
Since the convex hull is implied by the set of all bridges in $T$, we may still maintain the Partial Hull Tree with the previous root-to-leaf update strategy. 

\paragraph{Implicit bridges}
In a Partial Hull Tree $T$, the leaves store the values in $P_Y$, sorted by $x$-coordinate. 
I.e., we store the values of $Y$ in the leaves of $T$ in their stored order.
For a node $v$ with children $x$ and $y$, the bridge $e^+(v)$ is the bridge between the convex hulls $CH^+(\pi(x))$ and $CH^+(\pi(y))$. 
For a bridge $e^+(v)$, we can no longer store the endpoints of the bridge explicitly: as the $x$-coordinate of all bridges may radically change after an update in $Y$. 
We define the implicit bridge $\eps^+(v)$ which stores only the two values $(y_1, y_2)$ corresponding to the endpoints of $e^+(v)$. 
At this point, we wish to note that we can easily maintain the Partial Hull Tree  using implicit bridges with a factor $O(\log n)$ overhead.
Indeed, we may run any of the two proposed algorithms. 
Whenever we need to consider a bridge $\alpha = e^+(x)$, we can get the corresponding values $(y_1, y_2)$ from the implicit bridge.
Then, we may perform a binary search over $Y$ to obtain their corresponding ranks.
Thus, at $O(\log n)$ overhead, we always have explicit access to the endpoints of $\alpha$.

In the remainder of this section we show that we can cleverly navigate $T$ to avoid this overhead. 
To this end, we recall that for a vertex $v$ its median $med(v)$ was the leftmost child in the right subtree rooted at $v$. 
We define the \emph{widths} $\omega^+_1(v)$ (and $\omega^+_2(v)$) to be the rank-difference between $y_1$ and $med(v)$ (and $y_2$ and $med(v)$). 
Finally, we denote $\omega(e^+(x)) = \omega^+_1(v) + \omega^+_2(v)$. 
See Figure~\ref{fig:hullchange}(b) for an example. 
In our Partial Hull Tree, we now require that each $v \in T$ stores the size of its subtree, a pointer to the median of $v$ $med(v)$, and two \emph{implicit bridges} $\eps^+(v), \eps^-(v)$
 plus their widths $\omega^+_1(v), \omega^+_2(v), \omega^-_1(v), \omega^-_2(v)$. We note:

\begin{figure}[b]
\centering
\includegraphics[]{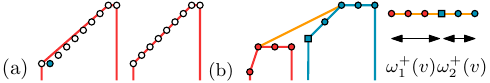}
\caption{
(a) A set $Y$ of values, mapped to $P_Y$. Deleting the blue value adds $O(n)$ edges to $CH(P_Y)$.
 (b) The bridge $e^+(v)$, the vertex $med(v)$ as a square and the corresponding widths.}
\label{fig:hullchange}
\end{figure}

\begin{lemma}
\label{lem:firstRank}
    Let $v \in T$, and denote by $r$ the rank of $med(v)$. 
    Given $r$ and $v$, we can compute the bridge $e^+(v)$ in $O(1)$ time. 
\end{lemma}

\begin{proof}
    The $x$-coordinate of the left endpoint of $e^+(v)$ is simply $r$ minus $\omega^+_1(v)$.
    The $x$-coordinate of the right endpoint of $e^+(v)$ is $r$ plus $\omega^+_1(v)$. 
The $y$-coordinates of both endpoints are stored in the implicit bridge $\eps^+(v)$. 
\end{proof}

\subsection{Updates using c-queues}

Recall that for any $v \in T$, $\mathbb{E}(v)$ was a balanced binary tree on the convex hull edges $CH^+(\pi(v))$ in their cyclical ordering. 
If $v$ had a child $x$, then the concatenable queue $\mathbb{E}^*(x)$ was defined as the balanced binary tree $\mathbb{E}(X) \backslash \mathbb{E}(v)$. 
We now define $\mathcal{E}(v)$ and $\mathcal{E}^*(x)$ analogously, where the inner nodes of the balanced binary trees store implicit edges bridges instead. 
Moreover, we demand that every $\alpha \in \mathcal{E}(x)$ stores the sum over all its descendants $\beta$ of $\omega(\beta)$. 
Note that these sums can be maintained at $O(\log n)$ time per insertion, deletion, split or join in the tree.
Moreover, for each tree rotation, these sums can be updated in $O(1)$ time. 
When storing the convex hull for rank-ordered data, we demand that each node $v \in T$ stores $\mathcal{E}^*(v)$; with the aforementioned sums in each node.

 Let us insert or delete a value $y \in Y$. 
 The update algorithm consists of three components which we will explain in order:
\begin{itemize}[noitemsep]
    \item an initialization over the root-to-leaf traversal in $T$ to $y$,
    \item a bubble-up update strategy, where we compute for each node $v$ on the leaf-to-root path the new bridge $e^+(v)$. We compute the bridge with its exact coordinates; from which we will derive $\eps^+(v)$ and the widths in $O(1)$ time. Finally, 
    \item a search algorithm to recompute for $v$ the bridge $e^+(v)$.
\end{itemize}

\paragraph{Traversing down.}
We start at the root $r$ where $\mathcal{E}^*(r) = \mathcal{E}(r)$. 
We do a root-to-leaf traversal to $y$.
For each node $v$ that we encounter, we store the rank of $med(v)$.
We assume that as we arrive at a node $v$, we have $\mathcal{E}(v)$ as a balanced binary tree.
Let $y \in \pi(x)$ for the left child $x$ of $v$. 
We invoke splitHull($x$, $\mathcal{E}(v)$), store $\mathcal{E}_R(v)$ in $v$.
We compute $\mathcal{E}(x)$ from the result of splitHull and recurse. 

\paragraph{Bubbling up.}
In the leaf $l$ that contains $y$, the bridge $e^+(l)$ equals $(y, y)$ and $med(l) = y$.  
We then consider each node $v$ on the leaf-to-root path from $l$ in $T$. 
Note that we have stored the rank $r$ of $med(v)$
Using $r$, $\mathcal{E}(x)$, and $\mathcal{E}_R(v)$, we recompute $e^+(v)$.
From $e^+(v)$ and $r$, we compute $\eps^+(v)$ and the corresponding widths.  
This allows us to compute $\mathcal{E}(v)$ and continue upwards in $T$.
If our update requires us to rotate $v$ in $T$, we recompute the bridge of all rotating nodes. 

\paragraph{Recomputing bridges.}
Given a node $v \in T$,  the rank $r$ of $med(v)$, $\mathcal{E}(x)$, and $\mathcal{E}_R(v)$ we do as follows. 
Denote by $x$ the left child of $v$ and by  $z$ the right child of $v$. 
From $r$, we can compute the ranks of $med(x)$ and $med(z)$ in $O(1)$ time. 
Given $\mathcal{E}_R(v)$ we can compute $\mathcal{E}(z)$ from $\mathcal{E}^*(z)$ using a single join operation. 
The root of $\mathcal{E}(x)$ is $\alpha' = \eps^+(x)$. We apply Lemma~\ref{lem:firstRank} to compute $\alpha = e^+(x)$ in $O(1)$ time.
Similarly, we obtain $\beta = e^+(z)$ in $O(1)$ time.
It follows that we may apply Lemma~\ref{lem:decision}. 

Suppose that the Lemma indicates that we may discard the subtree of $\mathbb{E}(x)$ right of $\alpha$. 
Let $\gamma = e^+(w)$ be the left child of $\alpha$ (for some $w \in T$). 
Denote by $\gamma' = \eps^+(w)$ the left child of $\gamma'$.
The rank of the right endpoint of $\gamma$ is equal to the rank of the left endpoint of $\alpha$, minus the width of all edges between $\gamma$ and $\alpha$.
Thus, we may compute the edge $\gamma$ from $\gamma'$ in $O(1)$ time by subtracting the width sum stored in $\gamma$. 
We can perform a symmetrical procedure to compute an edge $\zeta \in \mathbb{E}(z)$ and recurse to compute $e^+(v)$. 
Doing this for all $v \in T$ on the leaf-to-root path from the update implies the following theorem:

\begin{theorem}
    Let $Y$ be a dynamic set of values.  
    We can maintain the edges of $CH^+(P_Y)$ (stored as implicit edges $\eps(w)$) subject to insertions and deletions in $P$ as a balanced binary tree $\mathcal{E}(r)$ in $O(\log^2 n)$ worst-case time per update. 
\end{theorem}

What remains is to show that we can support queries in $O(\log h)$ time. 
Let $r$ be the root of the Partial Hull Tree $T$ and denote by $\mathcal{E}(r)$ the corresponding concatenable queue. 
The root of $\mathcal{E}(r)$ is the implicit bridge $\eps^+(r)$.
We may maintain the rank of $med(r)$ at no additional overhead. 
Thus by Lemma~\ref{lem:firstRank}, we may compute the edge $\alpha$ at the root of $\mathbb{E}(r)$ in $O(1)$ time. 
From here, we can perform queries in the same way that we find bridges: deciding whether to go into the left or right subtree of our current edge in $\mathcal{E}(r)$, and computing the explicit edge at the root of the new subtree in $O(1)$ time. 
It follows that we may answer Queries $(1)$-$(6)$ in $O(\log h)$ time (where $h$ is the number of edges on the convex hull of $P_Y$.

\subsection{The update algorithm without c-queues}
Let $v \in T$ and $r$ be the rank of $med(v)$. 
Finally let $x$ be any child of $v$. 
In the previous section we showed that we may compute the rank $r'$ of $med(x)$ from $r$ in $O(1)$ time by subtracting from (or adding to) $r$ the size of the other subtree of $v$. 
By Lemma~\ref{lem:firstRank}, we may use $r'$ to compute $e^+(x)$ from $\eps^+(x)$ in $O(1)$ time.
This implies that by maintaining our new concatenable queues $\mathcal{E}^*(x)$, we may run the algorithm of Section~\ref{sec:without}, without any additional overhead (other than maintaining for every node $v \in T$ the size of its subtree).

Our repository also includes implementations for maintaining rank-based convex hulls.

\section{Applying Dynamic Convex Hull}
\label{sec:applications}

We review three applications of a dynamic convex hull in VDLB. 

\paragraph{Query scheduling.}
For the query scheduling problem we consider incoming database queries, which each come with associated costs and gains. 
Efficient queries may be prioritized, whereas expensive queries may even be dropped if they bring too little gain compared to the other current queries.
The goal is to prioritize queries to maximize the gains. 
At VDLB 2011, Chi, Moon, and Hac{\'\i}g{\"u}m{\"u}{\c{s}}~\cite{chi2011icbs} study this problem, assuming that the cost/gain function of queries follows a piecewise linear function.  I.e., the gains of answering a query diminish piecewise linearly with the response time. 
They present a heuristic optimizer that relies upon a dynamic convex hull algorithm.
This convex hull algorithm  maps (dualizes) the piecewise linear functions to two-dimensional points.
Because they work with dualized points, they need to maintain the convex hull explicitly. 
They maintain the convex hull using the Overmars and van Leeuwen algorithm. 
They subsequently can use the convex hull to deduce which queries to prioritize. 
At VDLB 2013, Chi, Hac{\'\i}g{\"u}m{\"u}{\c{s}}, Hsiung, and Naughton~\cite{chi2013distribution} study the same problem, under the assumption that there exists some underlying cost distribution on query time per query type. 
On a high level, they design an equation that maps every query $q$ to a two-dimensional point which they call their Shepard score.
They subsequently want to prioritize queries with critical Shepard scores.
To this end, they maintain the Shepard scores (dualized) in the convex hull data structure by Overmars and van Leeuwen.
The authors present their own Java implementation to maintain this hull, which can be found at~\cite{javaimpl}.

\paragraph{Mining bursty subgraphs}
In temporal graphs, each edge can be represented as a triple $(u, v, t)$, where $u, v$ are two end nodes of one edge and $t$ denotes the
interaction time between $u$ and $v$~~\cite{qin2022mining}. 
A time-frame is \emph{bursty} whenever a large number of events occurs in a
a short time.
In VDLB 2022, Qin, Li, Yuan, Wang, Qin and Zhang~\cite{qin2022mining} study the problem of `mining' bursty patterns in a temporal graph. 
They present a dynamic program, that for every vertex $u$ in the graph computes the cumulative density of a `front' from $u$ in the graph $G$ over time. 
To compute this front, they repeatedly recompute the (lower) convex hull of their function from $u$. This step may be replaced by a dynamic convex hull algorithm.

\paragraph{PGM index}
Ferragina and Vinciguerra~\cite{ferragina2020pgm} study a dynamic sorted set of values $Y$ subject to rank queries.
They map $Y$ to the point set $P_Y$ and choose some $\eps >  0$. 
A segment $s$ $\eps$-\emph{covers} a consecutive set of values $(y_1, \ldots, y_k)$ whenever $s$ is within distance $\eps$ of all corresponding points $(p_1, \ldots, p_k)$. 
An $\eps$-cover is a set of segments $S$ that together $\eps$-cover all points in $P_Y$.
Denote by $m_\eps$ the minimal size for an $\eps$-cover. 
The authors present an amortized $O(\log n)$ algorithm to maintain an $\eps$-cover. 
They claim to support rank queries in $O(\log n (\log m_{\eps} + \log \eps))$ time. 
Their approach (under the hood) maintains the convex hull of $P_Y$ through the logarithmic method: splitting $Y$ over buckets $B$, and maintaining for the points $P_B$ a minimal $\eps$-cover (through indirectly maintaining $CH(P_B)$). 
Their approach has the same pitfall as hull queries under the logarithmic method: when $Y$ is split over buckets, the complexity $CH(P_B)$ (and thereby the complexity of the $\eps$-cover)  may significantly exceed that of $CH(P_Y)$. 
Without an algorithm to maintain $CH(P_Y)$ explicitly, their query time is $O(\log n (\log n + \log \eps))$. 

\section{The Logarithmic Method and Convex Hull}\label{app:logmethod}
Here we describe the logarithmic technique applied to the convex hull, and why doing so leads to incorrect query algorithms and hulls. 

\begin{figure}[h]
\centering
\includegraphics[]{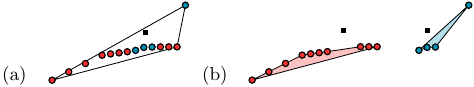}
\caption{(a) A point set $P$ and its convex hull.
(b) When we split $P$ over two buckets $B$, $B'$, the hulls $CH(\pi(B))$ and $CH(\pi(B'))$ have considerably higher complexity, and don't contain the black query square.
}
\label{fig:noconvex}
\end{figure}

Consider $P$ under insertions and Graham's scan algorithm (which uses $O(n)$ time for a sorted input set). 
We store $P$ in buckets $B_i$ of $2^i$ elements, where each $B_i$ is either full or empty.
Denote by $\pi(B) \subset P$ the points stored in a bucket $B$. For each bucket $B$ we store the convex hull $CH(\pi(B))$
Whenever we insert an element, we insert it in $B_2$. 
Suppose that an insertion causes $B_2$ to be full and let the first $k$ buckets be full. 
We then merge buckets $B_2, \ldots, B_k$ into the new bucket $B_{k+1}$.
During the merge, we ensure that the new bucket $B_{k+1}$ contains all values in sorted order in $O(2^{k+1})$ time by the same procedure as merge sort. 
Then, we construct the convex hull in $O( 2^{k+1})$  time using Graham's scan. This procedure has $O(\log n)$ amortized update time. However it does not maintain the convex hull of $P$, nor does it answer convex hull queries correctly: 

Consider (Figure~\ref{fig:noconvex}) a set $P$ of red and blue points where the convex hull of $P$ contains three points.
When we split the convex hull over two buckets (red and blue) both convex hulls contain many points.
Moreover, suppose that we query whether a point $q$ lies in $CH(P)$ (the black 
square).
It may be that $q$ is in none of the convex hulls $CH(B_i)$, even though it does lie in $CH(P)$.  
This inconsistency  causes~\cite{ferragina2020pgm} to store something slightly different from advertised (Appendix~\ref{sec:applications}).

\section{Additional Results}\label{app:results}
\begin{figure}[htb]
\centering
\includegraphics[width=.99\linewidth]{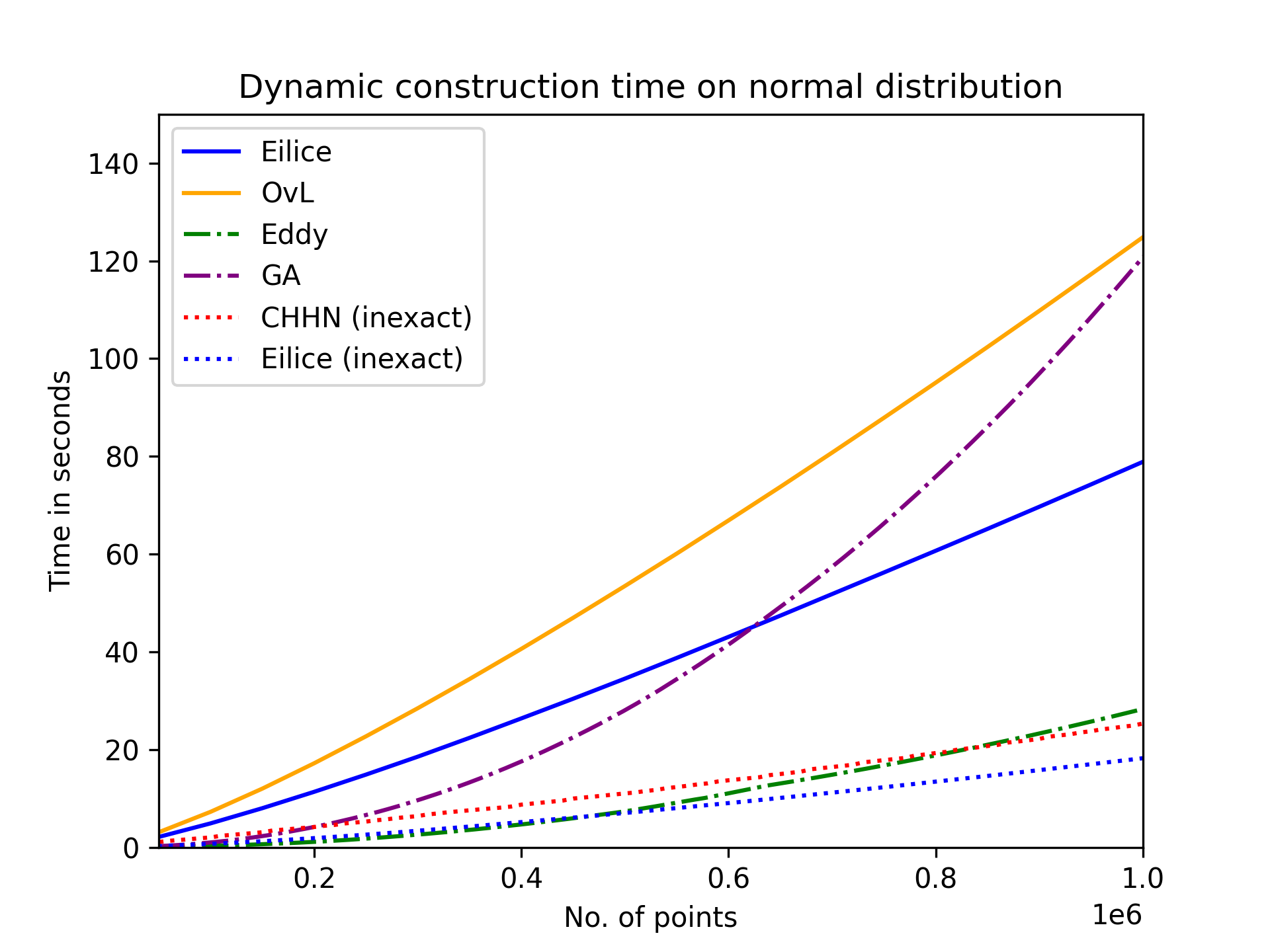}
\includegraphics[width=.99\linewidth]{Figures/Results/construct_unif.png}

\caption{
    Dynamic construction time using $500$ point extensions on various distributions.
    }
\label{fig:construct_total}
\end{figure}
\begin{figure}[htb]
\centering
\includegraphics[width=.99\linewidth]{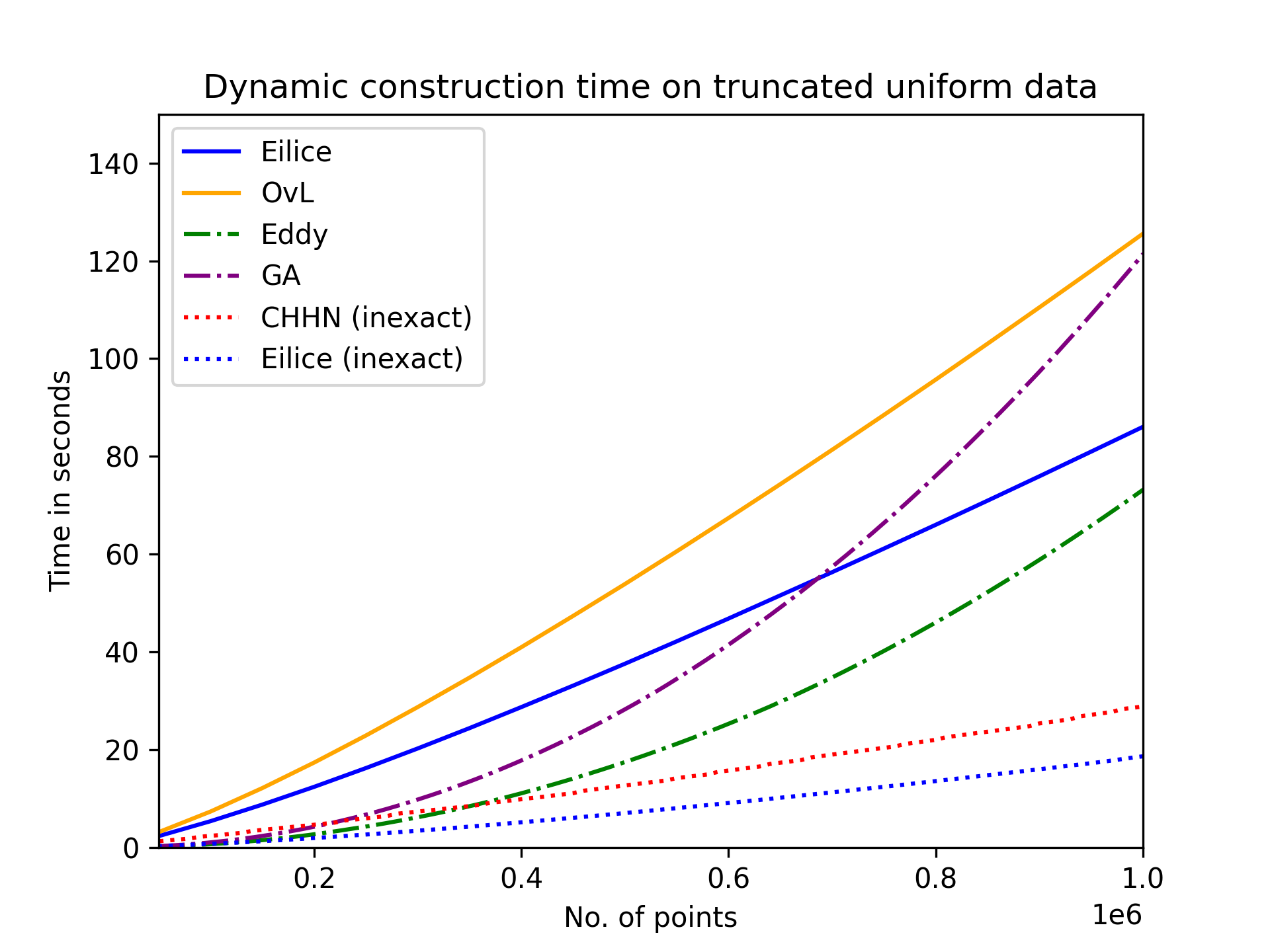}
\includegraphics[width=.99\linewidth]{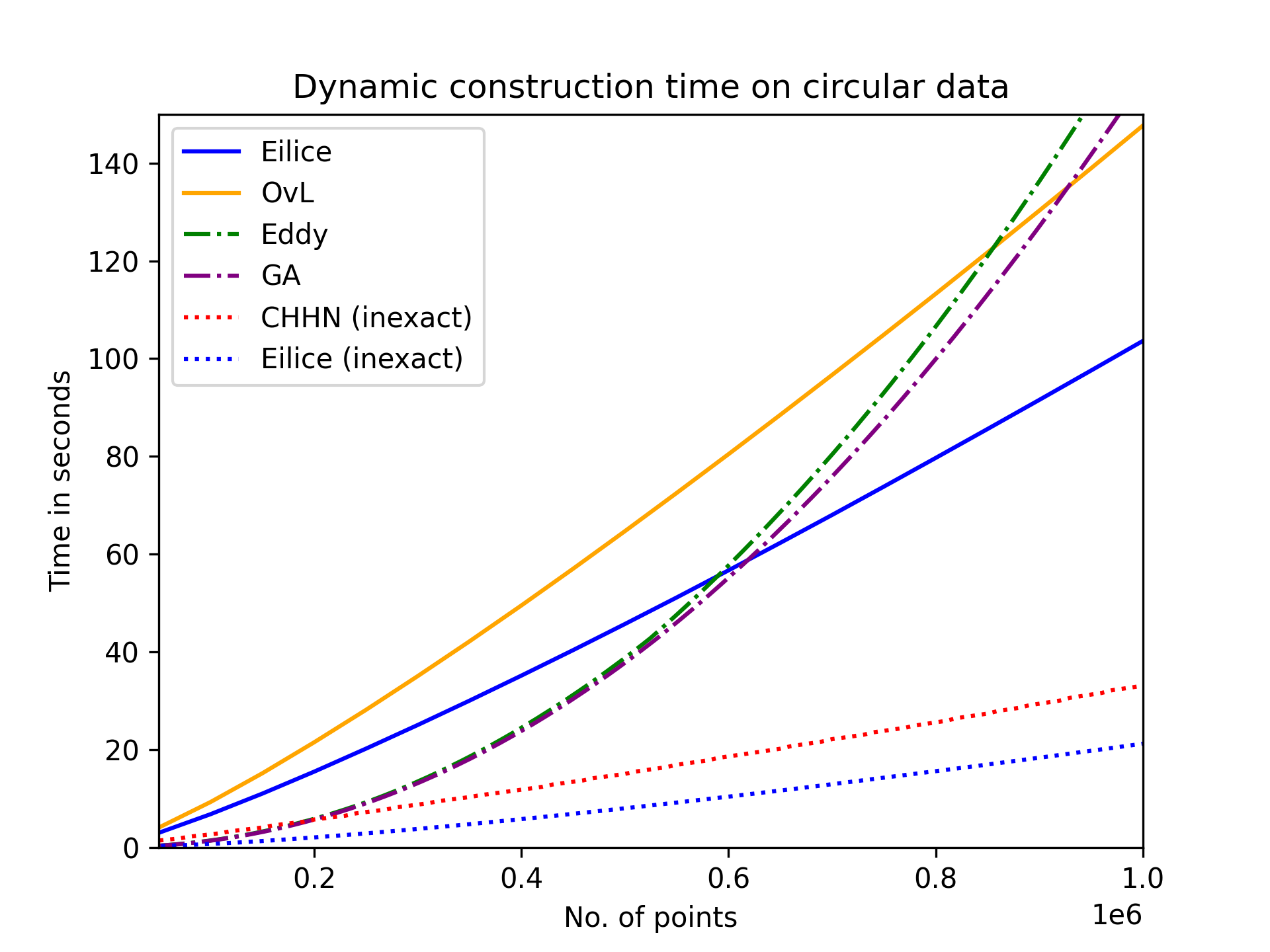}

\caption{
    Dynamic construction time using $500$ point extensions on various distributions.
    }
\label{fig:construct_total}
\end{figure}
\begin{figure}[htb]
\centering
\includegraphics[width=.99\linewidth]{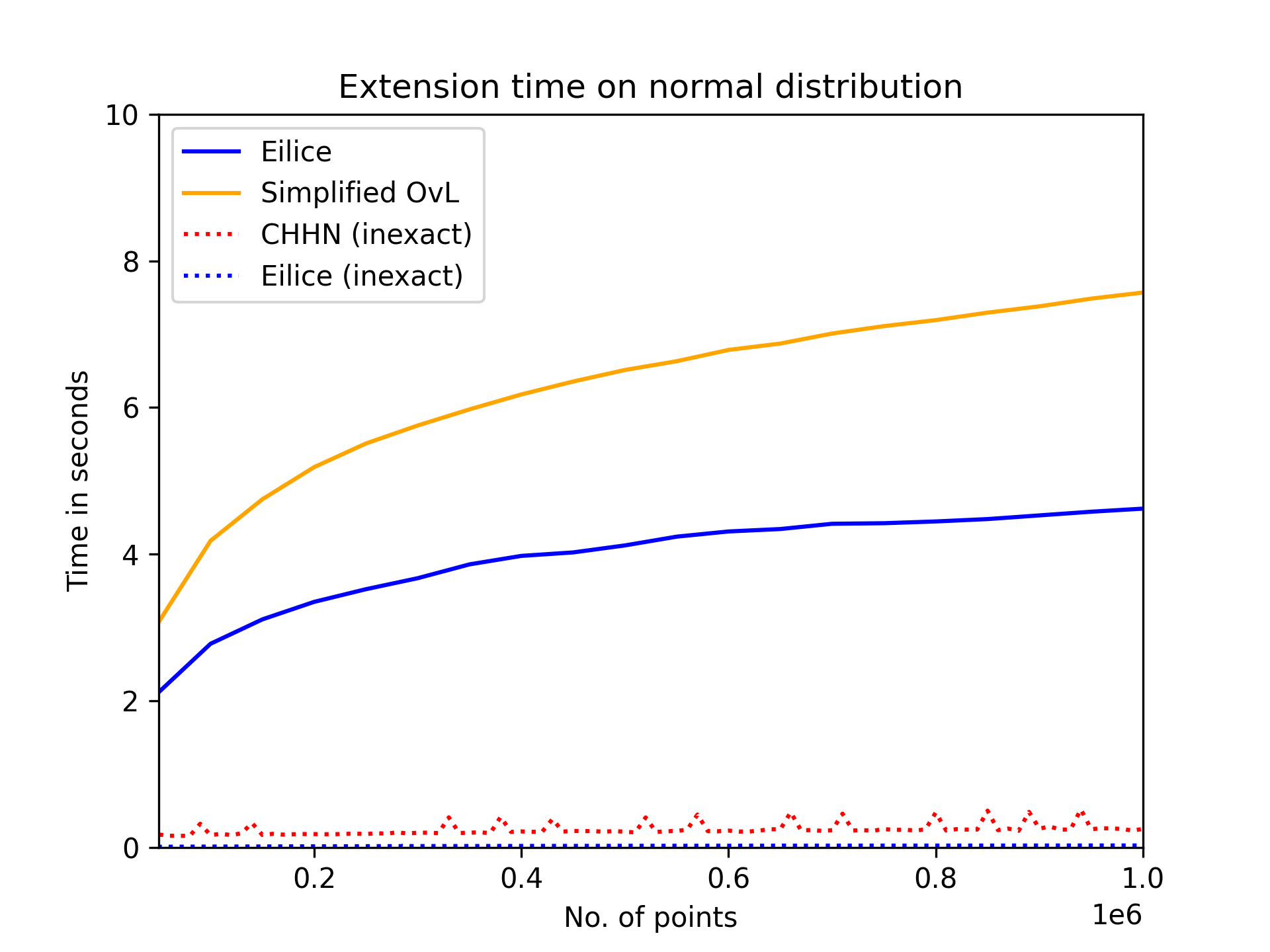}
\includegraphics[width=.99\linewidth]{Figures/Results/extend_unif.png}

\caption{
    Extension time on various distributions.
    }
\label{fig:extend_total}
\end{figure}
\begin{figure}[htb]
\centering
\includegraphics[width=.99\linewidth]{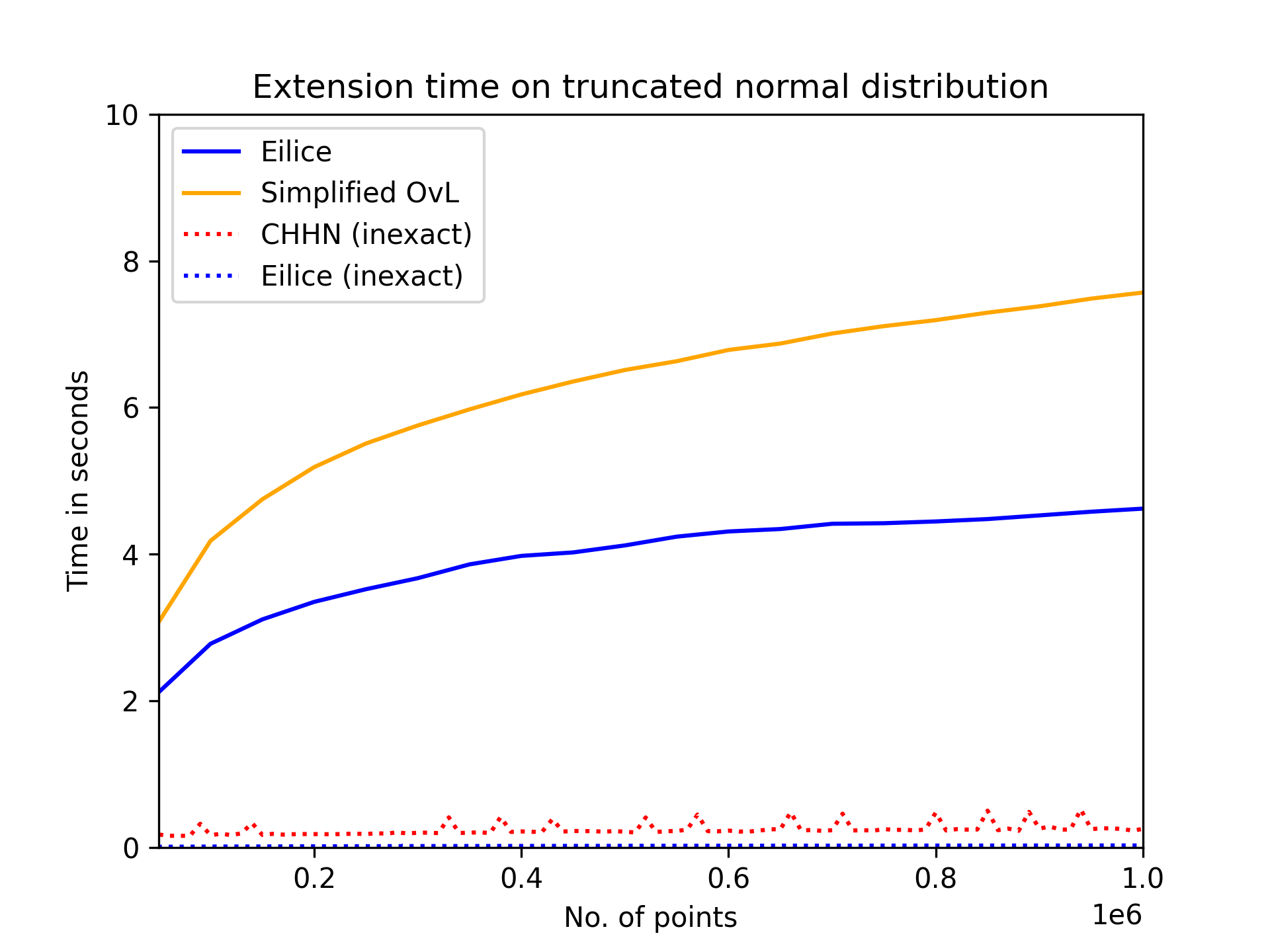}
\includegraphics[width=.99\linewidth]{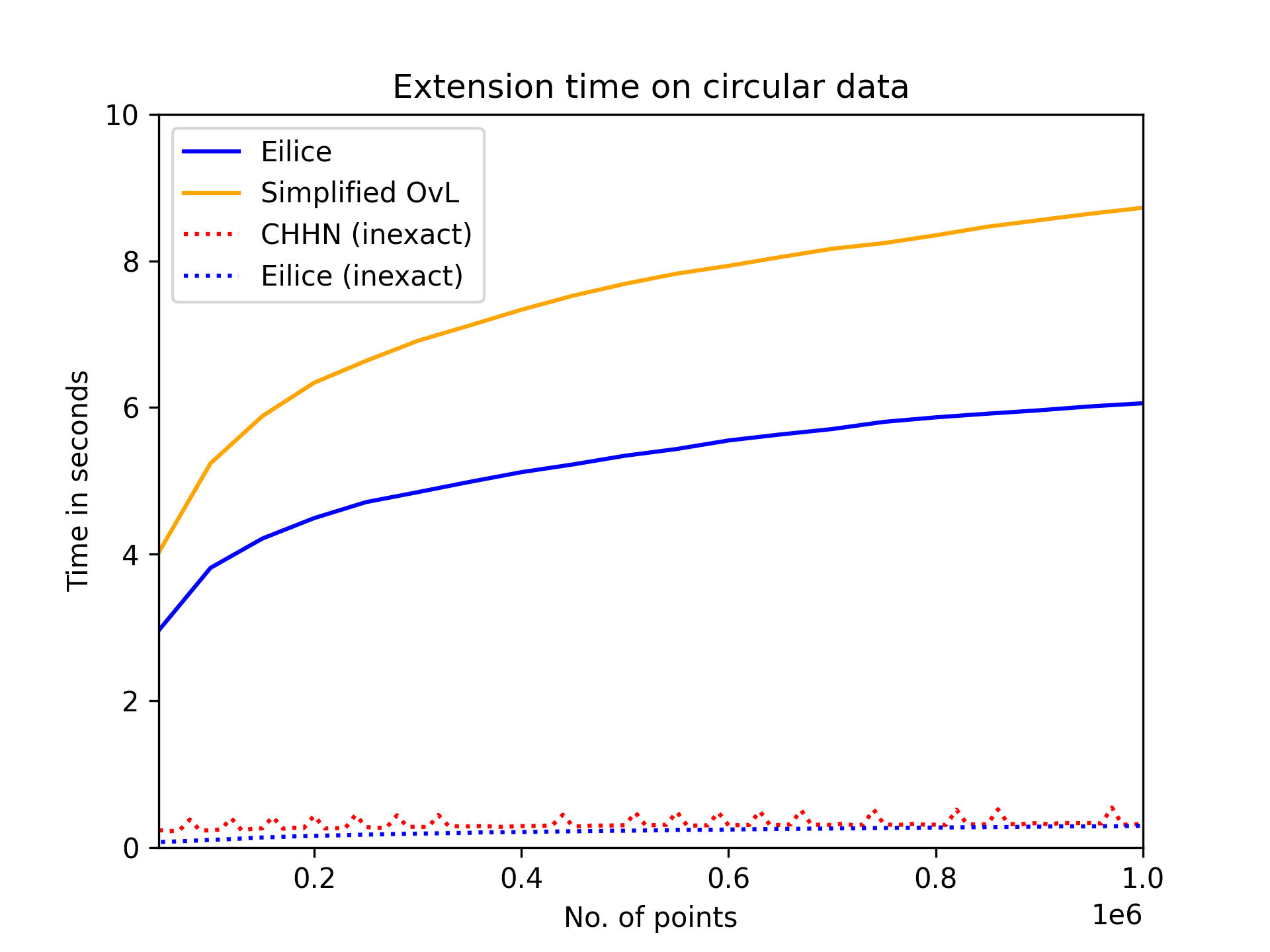}

\caption{
    Extension time on various distributions.
    }
\label{fig:extend_total}
\end{figure}
\begin{figure}[htb]
\centering
\includegraphics[width=.99\linewidth]{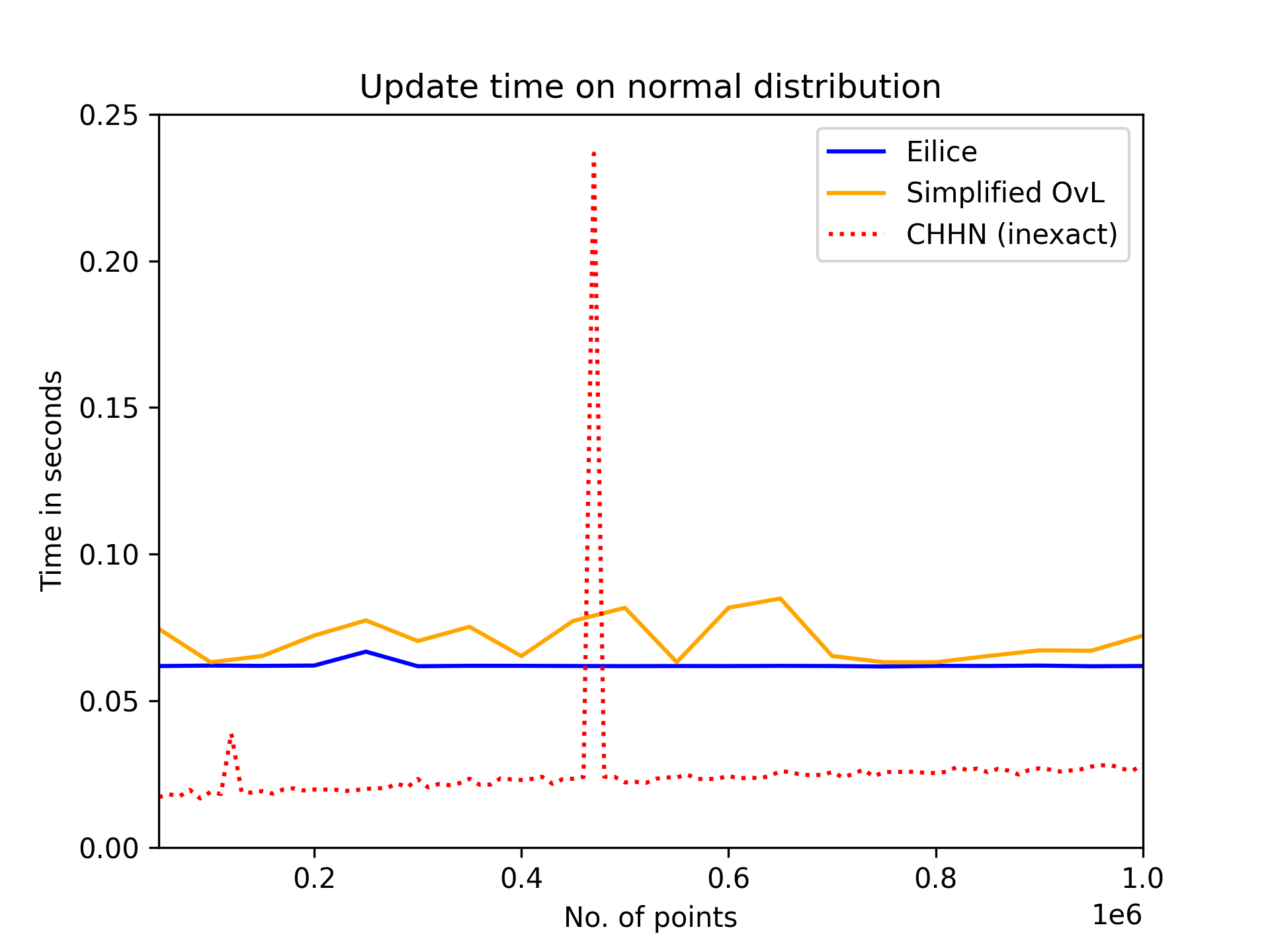}
\includegraphics[width=.99\linewidth]{Figures/Results/update_unif.png}

\caption{
    Update time on various distributions.
    }
\label{fig:update_total}
\end{figure}
\begin{figure}[htb]
\centering
\includegraphics[width=.99\linewidth]{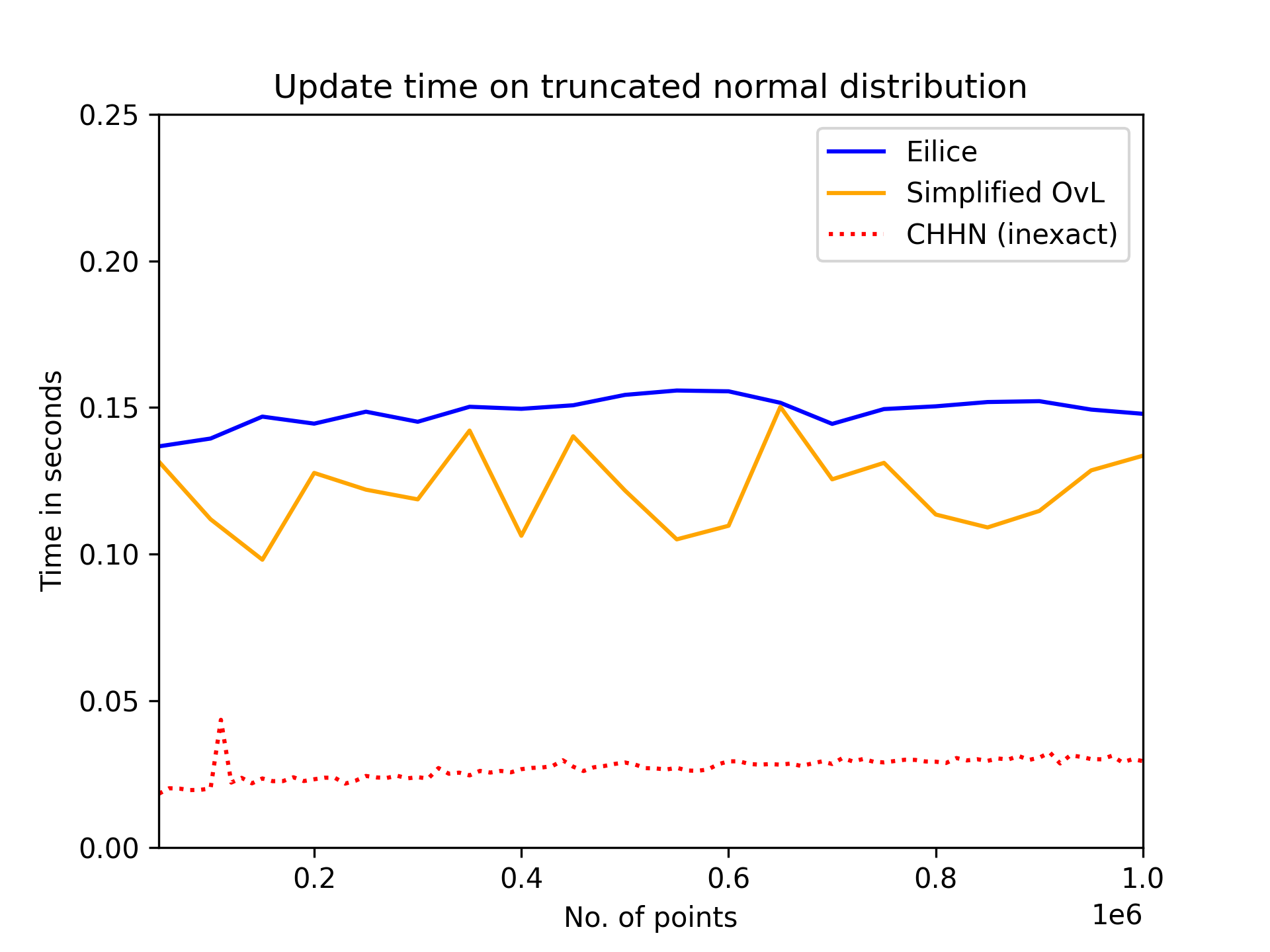}
\includegraphics[width=.99\linewidth]{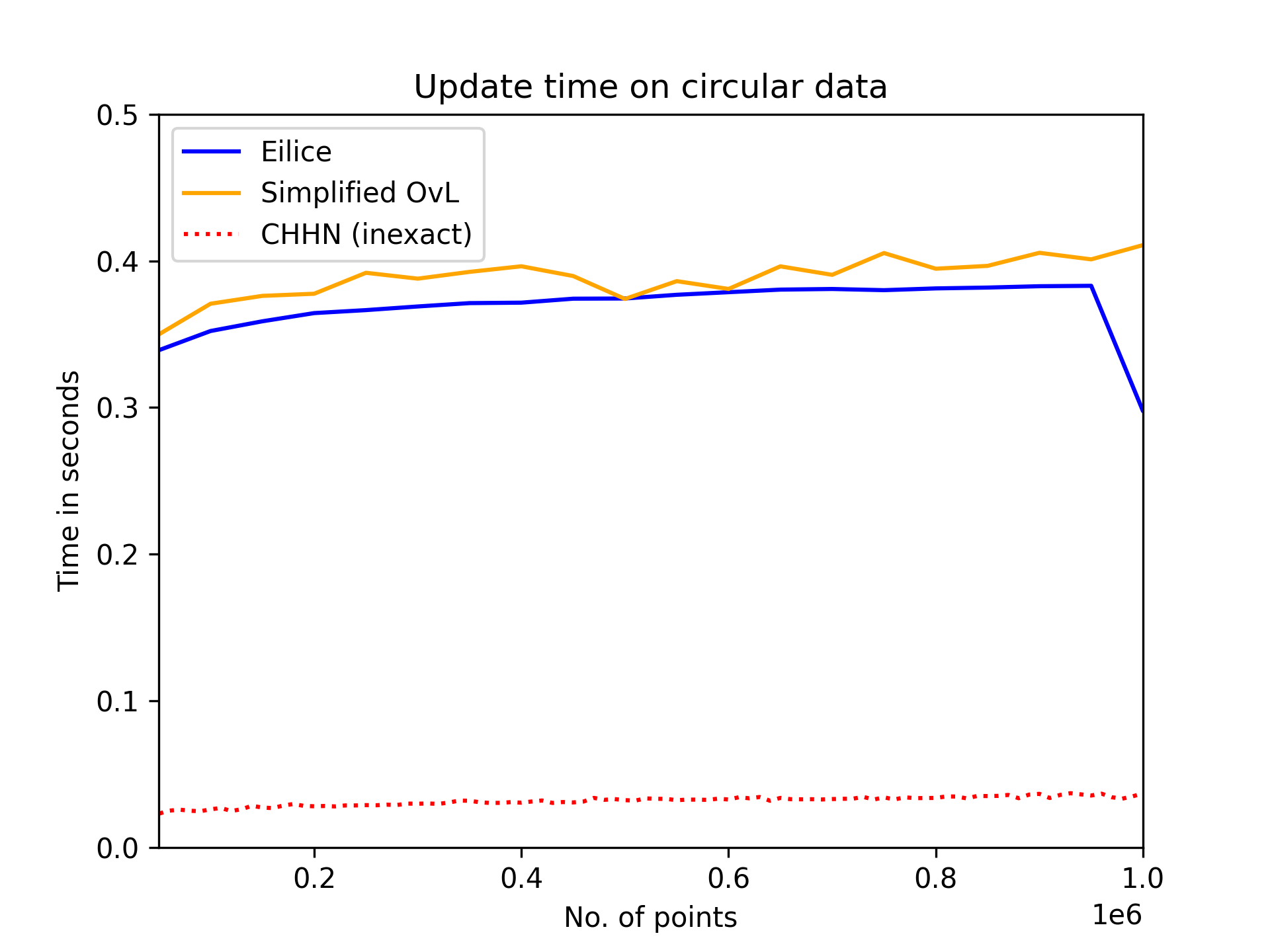}

\caption{
    Update time on various distributions.
    }
\label{fig:update_total}
\end{figure}
\begin{figure}[htb]
\centering
\includegraphics[width=.99\linewidth]{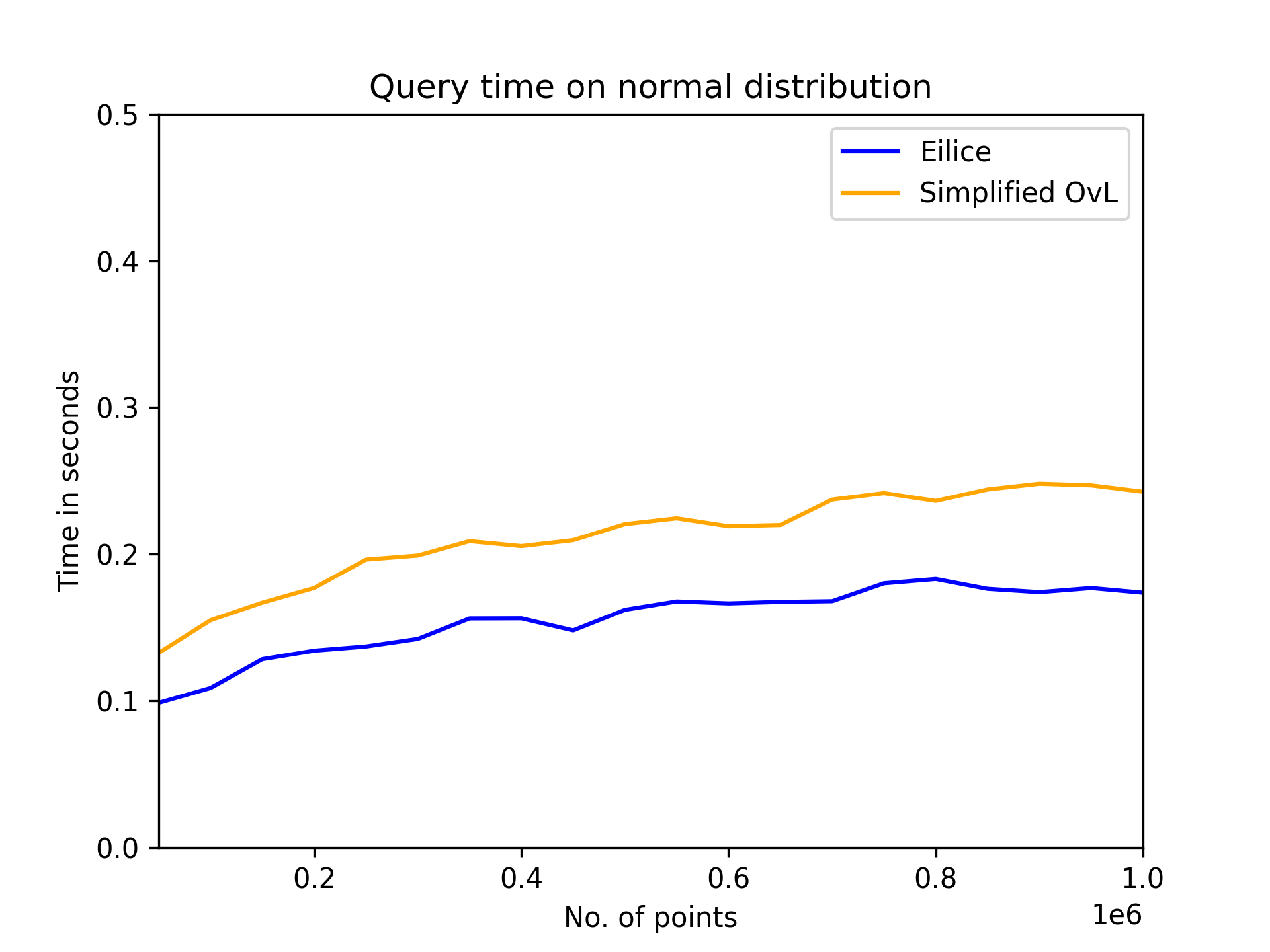}
\includegraphics[width=.99\linewidth]{Figures/Results/query_unif.png}

\caption{
    Query time on various distributions.
    }
\label{fig:query_total}
\end{figure}
\begin{figure}[htb]
\centering
\includegraphics[width=.99\linewidth]{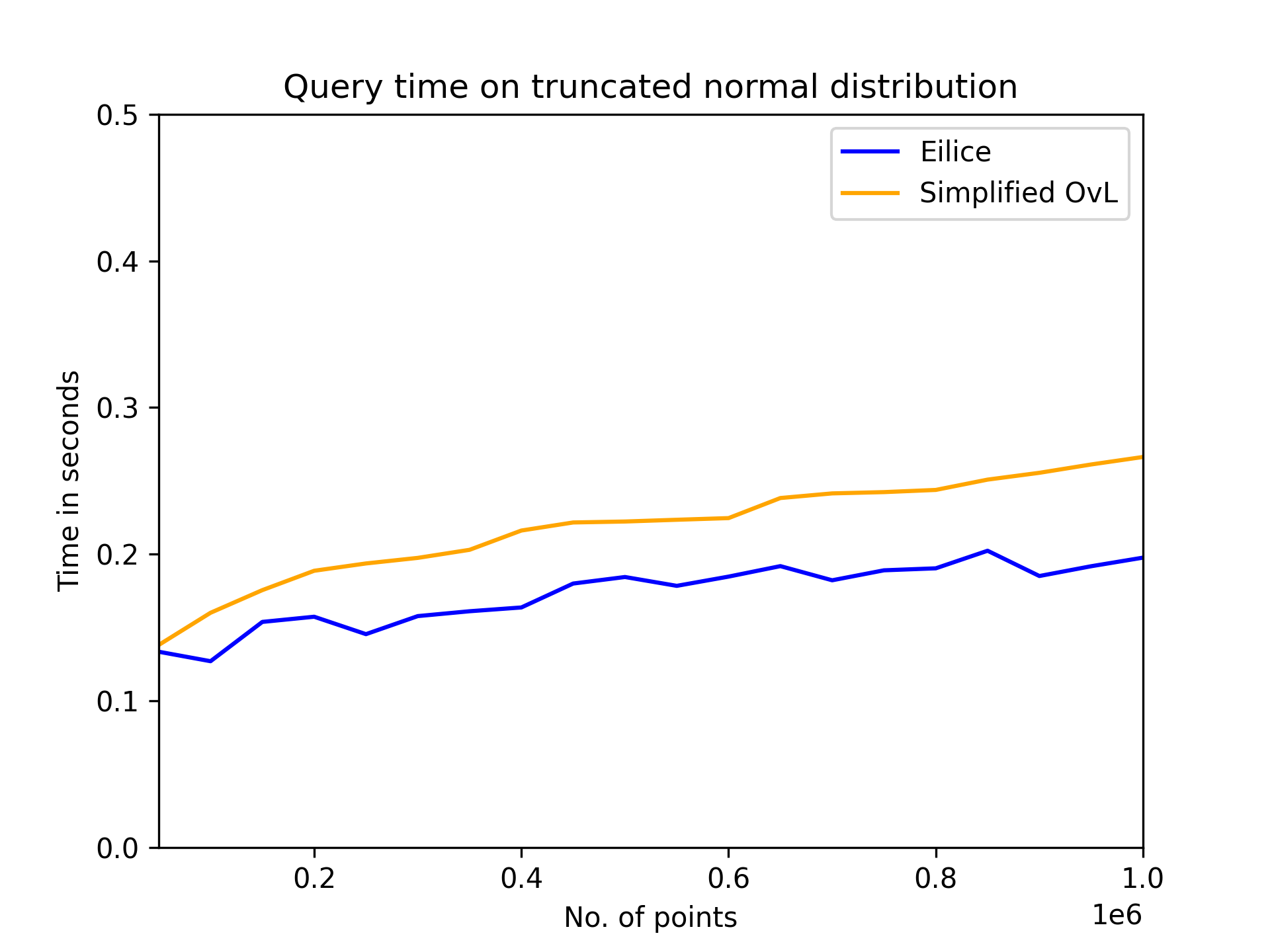}
\includegraphics[width=.99\linewidth]{Figures/Results/query_circ.png}

\caption{
    Query time on various distributions.
    }
\label{fig:query_total}
\end{figure}
\end{document}